\documentclass[12pt]{article}

\pdfoutput=1 %

\usepackage{jheppub} %
\usepackage{graphicx}  %
\usepackage{dcolumn}   %
\usepackage{bm}        %
\usepackage{amssymb}   %
\usepackage{subfigure}
\usepackage{epstopdf}
\usepackage{color}
\usepackage{slashed}
\usepackage[utf8]{inputenc}
\usepackage{multirow}
\usepackage{footnote}
\usepackage{amsmath}
\allowdisplaybreaks[4]
\usepackage{accents}
\usepackage{gensymb}
\usepackage{color}

\newlength{\dhatheight}

\def\figureautorefname~#1\null{Fig.\,#1\null}
\def\tableautorefname~#1\null{Tab.\,#1\null}

\def\equationautorefname~#1\null{Eq.\,(#1)\null}

\graphicspath{{figs/}}

\title{Search for Light Neutral Scalar in the Georgi-Machacek Model with Forward Detectors at the LHC}
\author[a,b]{Yuanqian Fang,}
\author[c]{Wei Su,}
\author[a,b]{Xinyu Wang,}
\author[a,b]{Yongcheng Wu,}
\author[a,b]{Yan Zhang}
\affiliation[a]{Department of Physics and Institute of Theoretical Physics, Nanjing Normal University, Nanjing, 210023, China}
\affiliation[b]{Nanjing Key Laboratory of Particle Physics and Astrophysics, Nanjing Normal University, Nanjing, 210023, China}
\affiliation[c]{School of Science, Shenzhen Campus of Sun Yat-sen University, No. 66, Gongchang Road, \\ Guangming District, Shenzhen, Guangdong 518107, China}
\emailAdd{yqfang@njnu.edu.cn}
\emailAdd{suwei26@mail.sysu.edu.cn}
\emailAdd{xinyuwang@njnu.edu.cn}
\emailAdd{ycwu@njnu.edu.cn}
\emailAdd{zyan@njnu.edu.cn}
\preprint{CPTNP-2025-046}

\abstract{
Long-lived particle (LLP) is one of the well-motivated targets for current collider experiments searching for the physics beyond the Standard Model. In recent years, many dedicated detectors have been developed for such scenarios which are designed to extend the sensitivity to weakly coupled particles with macroscopic $c\tau$. In this work, we investigate the LLP signatures of the neutral component of the fermiophobic fiveplet $H_5^0$ in the Georgi-Machacek model. Due to its fermiophobic  nature at tree level, it possesses suppressed decay widths in the low mass region and can naturally be long-lived over a wide region of parameter space. We show that $H_5^0$ can be produced with an appreciable flux in the forward region from the meson decay through loop-induced couplings leading to observable LLP signatures in forward detectors. 
We evaluate the sensitivity of representative forward detectors to this scenario and compare the result with existing constraints from terrestrial experiments and astrophysical observations. Our results demonstrate that the forward detectors can probe the $s_H$ down to $\mathcal{O}(10^{-5})$ for sub-GeV scalar masses, and hence providing a powerful and complementary probe of the extended Higgs sectors that is inaccessible to conventional searches.
}

\begin{document}
\titlepage
\maketitle
\newpage

\flushbottom

\section{Introduction}

Since the discovery of the Higgs boson at the LHC in 2012~\cite{ATLAS:2012yve,CMS:2012qbp}, extensive searches for physics beyond the Standard Model (BSM) have been carried out by ATLAS and CMS. These searches target a wide variety of scenarios, including supersymmetry (SUSY), extended gauge sectors, and extra scalars. However, no statistically significant deviation from the Standard Model (SM) has been observed. As a consequence, stringent constraints have been placed on the mass and couplings of new states, even pushing the scale of BSM into several TeV regime when prompt decays are assumed for the analysis.

However, a broad class of theoretically and phenomenologically well motivated models continues to motivate the searches of new physics around electroweak scale. These include the explanations of neutrino masses~\cite{Mohapatra:2006gs}, the origin of baryon asymmetry of the Universe (BAU)~\cite{Dine:2003ax} and the dark matter~\cite{Bertone:2004pz}. Note that, the current null result from conventional searches does not exclude completely the possibility that there exist new light states if the couplings with SM particles are weak or if the decay width are highly suppressed. In such scenario, new particles can be long-lived on the detector scale and escape conventional searches based on prompt decays~\cite{Alimena:2019zri}. This has motivated increasing interests in long-lived particles (LLPs) as another possibility for BSM at the LHC. ATLAS and CMS have performed searches for displaced vertices, displaced jets, disappearing tracks constraining the decay length of LLP from millimeter to several meters~\cite{ATLAS:2018tup,CMS:2018qxv,ATLAS:2020wjh,CMS:2018rea,CMS:2019zxa,ATLAS:2024qoo,ATLAS:2017tny,ATLAS:2021jig,ATLAS:2024ocv,CMS:2024qxz,CMS:2023mny,CMS:2024trg,CMS:2024bvl,CMS:2021yhb,CMS:2021juv}.

Despite the progress, significant regions of LLP parameter space remain unexplored. In particular, neutral LLP with light mass and decay length exceeding the scale of the detectors of ATLAS and CMS are only weakly constrained. Such particles may decay after it escape the detectors of ATLAS and CMS resulting in a dramatic loss of the sensitivity. To address such gap, several dedicated LLP detectors have been proposed and developed, including FASER/FASER2~\cite{FASER:2018eoc}, MATHUSLA~\cite{Okada:2024kni,MATHUSLA:2022sze,Chou:2016lxi,Curtin:2018mvb,MATHUSLA:2020uve}, CODEX-b~\cite{Gligorov:2017nwh}, ANUBIS~\cite{Shah:2024fpl,Shah:2024tgd}, MoEDAL-MAPP1/MAPP2~\cite{MoEDAL-MAPP:2022kyr,Mitsou:2024nuf,Kalliokoski:2024kxv,MoEDAL-MAPP:2022kyr}, FACET~\cite{Cerci:2021nlb}, AL3X~\cite{Gligorov:2018vkc}. These experiments are designed to explore the signal of such weakly interacting particles produced from the interacting point (IP) and decaying in the far/forward region which can achieve exceptional sensitivity to light LLPs with long lifetimes. With FASER~\cite{FASER:2018eoc} already operational within the Forward Physics Facility (FPF)~\cite{Anchordoqui:2021ghd,Feng:2022inv} and others under active development, forward physics has become a central component of the LHC operations.

In this work, we focus on the LLP signatures arising in the Georgi-Machacek (GM) model, which extends the SM scalar sector while preserves the custodial symmetry at tree level~\cite{Georgi:1985nv,Chanowitz:1985ug,Gunion:1989ci,Hartling:2014zca,Aoki:2007ah}. The GM model introduces two extra $SU(2)_L$ triplets, one complex one real, leading to an enlarged scalar sectors which can be organized into custodial multiplets: two singlets (one of which will be identified as the 125 GeV Higgs), two triplets (one of which is the Goldstone) and one fiveplet. A particularly distinctive feature of the GM model is the scalar fiveplet, which is fermiophobic. They do not couple to the SM fermions at tree level. As a result, the decay widths of the fiveplets are controlled entirely by gauge interactions and scalar self-couplings. For the neutral component, when its mass is much smaller than the gauge bosons and other scalars, the decay width will be highly suppressed. In such region, the loop-induced Yukawa couplings play a crucial role for the searches of this particle. The light scalar can be produced through meson decays with these loop-induced Yukawa couplings which provides sufficient scalar flux in the forward region thanks to the extreme large flux of the mesons produced in this region. The forward detectors at the LHC thus provide an unique probe and are complementary to other searches at the LHC for the fermiophobic scalars in the GM model.

The rest of this paper is arranged as follows. In Sec.~\ref{Sec:GM_model}, we briefly introduce the GM model and relevant theoretical limitations. After that, the loop-induced Yukawa couplings for fermiophobic fiveplet is calculated in a general way. 
The productions and decays of the fiveplet are discussed in Sec.~\ref{Sec:Production_decay_H50}. After a brief discussion about current constraints from terrestrial and astrophysical observations, the results from forward detectors are presented for benchmark parameter spaces in Sec.~\ref{Sec:H50_analysis}. Then we summarize in Sec.~\ref{Sec:Conclusion}. The couplings that are relevant for the loop calculations are listed in Appendix~\ref{app:couplings}.

\section{The Georgi-Machacek Model}
\label{Sec:GM_model}
\subsection{The model setup}
In the GM model~\cite{Georgi:1985nv,Chanowitz:1985ug}, along with the standard scalar doublet, we have two extra triplets. In order to explicitly show the global ${\rm SU}(2)_L\times {\rm SU(2)}_R$ symmetry, these fields are presented in the follow bi-doublet and bi-triplet form:
\begin{align}
\Phi &= \begin{pmatrix}
    \phi^{0*} & \phi^+ \\
    -\phi^{+*} & \phi^0
\end{pmatrix},\quad X = \begin{pmatrix}
    \chi^{0*} & \xi^+ & \chi^{++} \\
    -\chi^{+*} & \xi^0 & \chi^+ \\
    \chi^{++*} & -\xi^{+*} & \chi^0
\end{pmatrix}
\end{align}
The most general gauge-invariant scalar potential conserving custodial symmetry~\cite{Hartling:2014zca} is given by
\begin{align}
    \label{equ:V0}
    V_0 &= \frac{\mu_2^2}{2}{\rm Tr}(\Phi^\dagger\Phi)+\frac{\mu_3^2}{2}{\rm Tr}(X^\dagger X) + \lambda_1\left({\rm Tr}(\Phi^\dagger\Phi)\right)^2 + \lambda_2{\rm Tr}(\Phi^\dagger\Phi){\rm Tr}(X^\dagger X) \nonumber \\
    & \qquad + \lambda_3{\rm Tr}(X^\dagger X X^\dagger X) + \lambda_4\left({\rm Tr}(X^\dagger X)\right)^2 - \lambda_5{\rm Tr}(\Phi^\dagger \tau^a\Phi\tau^b){\rm Tr}(X^\dagger t^a X t^b) \nonumber \\
    &\qquad- M_1 {\rm Tr}(\Phi^\dagger \tau^a \Phi \tau^b)(UXU^\dagger)_{ab} - M_2 {\rm Tr}(X^\dagger t^a X t^b)(UXU^\dagger)_{ab}
\end{align}
where $\tau^a=\sigma^a/2$ with $\sigma^a$ being the Pauli matrices. The generators for the triplet representation are given as
\begin{align}
    t^1 = \frac{1}{\sqrt{2}}\begin{pmatrix}
        0 & 1 & 0 \\
        1 & 0 & 1 \\
        0 & 1 & 0
    \end{pmatrix}, \quad t^2=\frac{1}{\sqrt{2}}\begin{pmatrix}
        0 & -i & 0 \\
        i & 0 & -i \\
        0 & i & 0
    \end{pmatrix}, \quad t^3=\begin{pmatrix}
        1 & 0 & 0 \\
        0 & 0 & 0 \\
        0 & 0 & -1
    \end{pmatrix}.
\end{align}
The matrix $U$ rotates $X$ into the Cartesian basis~\cite{Aoki:2007ah} and is given by
\begin{align}
    U = \begin{pmatrix}
        -\frac{1}{\sqrt{2}} & 0 & \frac{1}{\sqrt{2}}\\
        -\frac{i}{\sqrt{2}} & 0 & -\frac{i}{\sqrt{2}}\\
        0 & 1 & 0
    \end{pmatrix}.
\end{align}
After the electroweak symmetry breaking (EWSB), the scalar fields acquire vevs as
\begin{align}
\label{equ:vev}
    \langle\Phi\rangle = \frac{v_\phi}{\sqrt{2}}\mathbb{I}_{2\times2},\quad \langle X\rangle = v_\chi\mathbb{I}_{3\times3},
\end{align}
where the vevs satisfy $v_\phi^2 + 8v_\chi^2=v^2\approx(246\,{\rm GeV})^2$. 
We introduce the angle $\theta_H$ defined from the vevs as
\begin{align}
    c_H\equiv \cos\theta_H = \frac{v_\phi}{v},\quad s_H\equiv\sin\theta_H = \frac{2\sqrt{2}v_\chi}{v}.
\end{align}
In terms of the vevs, the scalar potential is given by
\begin{align}
\label{equ:Vvev}
V(v_\phi,v_\chi) &= \frac{1}{2}\mu_2^2v_\phi^2+\frac{3}{2}\mu_3^2v_\chi^2 + \lambda_1 v_\phi^4 + \frac{3}{2}(2\lambda_2-\lambda_5)v_\phi^2v_\chi^2 + 3(\lambda_3 + 3\lambda_4)v_\chi^4\nonumber \\
&\qquad\qquad - \frac{3}{4}M_1v_\phi^2v_\chi - 6M_2v_\chi^3.
\end{align}
The vevs are then determined by the minimum position of~\autoref{equ:Vvev}:
\begin{align}
    \label{equ:stationary_condition}
    \begin{cases}
    \frac{\partial V}{\partial v_\phi} = 0, \\
    \frac{\partial V}{\partial v_\chi} = 0,
    \end{cases} \quad \Rightarrow\quad \begin{cases}
    \mu_2^2 = -4\lambda_1v_\phi^2+3(\lambda_5-2\lambda_2)v_\chi^2 + \frac{3}{2}M_1v_\chi,\\
    \mu_3^2 = (\lambda_5-2\lambda_2)v_\phi^2-4(\lambda_3+3\lambda_4)v_\chi^2+\frac{M_1v_\phi^2}{4v_\chi}+6M_2v_\chi.
    \end{cases}
\end{align}

The physical fields will be fluctuations around the vacuum, and we define for the neutral components
\begin{align}
    \phi^0 = \frac{v_\phi}{\sqrt{2}} + \frac{\phi^{0,r}+i\phi^{0,i}}{\sqrt{2}}, \quad \chi^0 = v_\chi + \frac{\chi^{0,r}+i\chi^{0,i}}{\sqrt{2}}, \quad \xi^0 = v_\chi + \xi^0.
\end{align}
The mass of the scalars are determined by the quadratic terms of the potential. After electroweak symmetry breaking, the custodial symmetry is preserved. Hence, the scalars can be grouped into multiplets under the custodial symmetry:
\begin{subequations}
    \begin{align}
        H_5^{\pm\pm} &= \chi^{\pm\pm},\\
        H_5^\pm &= \frac{\chi^\pm-\xi^\pm}{\sqrt{2}},\\
        H_5^0 &= \sqrt{\frac{2}{3}}\xi^0 - \sqrt{\frac{1}{3}}\chi^{0,r},\\
        H_3^\pm &= - s_H \phi^\pm + c_H \frac{\chi^\pm+\xi^\pm}{\sqrt{2}},\\
        H_3^0 &= -s_H \phi^{0,i} + c_H \chi^{0,i},\\
        G^\pm &= c_H\phi^\pm + s_H\frac{\chi^\pm+\xi^\pm}{\sqrt{2}},\\
        G^0 &= c_H \phi^{0,i} + s_H \chi^{0,i},\\
        H_1^0 &= \phi^{0,r}, \\
        H_1^{0\prime} &= \sqrt{\frac{1}{3}}\xi^0 + \sqrt{\frac{2}{3}}\chi^{0,r}.
    \end{align}
\end{subequations}
The masses for the fiveplet and triplets are
\begin{align}
    m_5^2 &= \frac{M_1}{4v_\chi}v_\phi^2 + 12M_2v_\chi + \frac{3}{2}\lambda_5 v_\phi^2 + 8\lambda_3 v_\chi^2,\label{equ:m5_mass}\\
    m_3^2 &= \frac{M_1}{4v_\chi}(v_\phi^2+8v_\chi^2)+\frac{1}{2}\lambda_5 (v_\phi^2 + 8 v_\chi^2) = \left(\frac{M_1}{4v_\chi}+\frac{\lambda_5}{2}\right)v^2.
\end{align}
The singlets listed above $(H_1^0, H_1^{0\prime})$ will further mix with mixing angle $\alpha$ to form the mass eigenstates $(h,H)$:
\begin{align}
\label{equ:mixing_singlet}
\begin{pmatrix}
    h\\
    H
\end{pmatrix} = \begin{pmatrix}
    c_\alpha & -s_\alpha \\
    s_\alpha & c_\alpha
\end{pmatrix}\begin{pmatrix}
    H_1^0\\
    H_1^{0\prime}
\end{pmatrix}.
\end{align}
The mass matrix in the $(H_1^0, H_1^{0\prime})$ basis is
\begin{align}
\label{equ:mass_singlet}
\mathcal{M}^2 = \begin{pmatrix}
    8\lambda_1v_\phi^2 & \frac{\sqrt{3}}{2}v_\phi\left(-M_1+4(2\lambda_2 - \lambda_5)v_\chi\right)\\
    \frac{\sqrt{3}}{2}v_\phi\left(-M_1+4(2\lambda_2 - \lambda_5)v_\chi\right) & \frac{M_1v_\phi^2}{4v_\chi}-6M_2v_\chi+8(\lambda_3+3\lambda_4)v_\chi^2
\end{pmatrix}.
\end{align}
The masses for two singlets are 
\begin{align}
\label{equ:mh_mH}
	m^2_{h,H} = \frac{1}{2} \left[ \mathcal{M}_{11}^2 + \mathcal{M}_{22}^2
	\mp \sqrt{\left( \mathcal{M}_{11}^2 - \mathcal{M}_{22}^2 \right)^2 
	+ 4 \left( \mathcal{M}_{12}^2 \right)^2} \right].
\end{align}
The scalar potential~\autoref{equ:V0} has 9 free parameters, including the dimensional $\mu_2^2,\mu_3^2,M_1,M_2$ and dimensionless $\lambda_i's$. 
Using~\autoref{equ:stationary_condition}, \autoref{equ:m5_mass} and \autoref{equ:mh_mH}, the $\mu_2^2$, $\mu_3^2$, $M_1$ and $\lambda_1$ can be calculated from $v$, $s_H$, $m_h$ and $m_5$ and other parameters in the potential. Hence, in this work, the following parameters are used as inputs
\begin{align}
v(\approx246\,{\rm GeV}),\, s_H,\, m_h(\approx125\,{\rm GeV}),\, m_5,\, \lambda_2,\, \lambda_3,\, \lambda_4,\, \lambda_5,\, M_2,
\end{align}
which is aligned with the low-$m_5$ benchmark used in~\cite{Ismail:2020kqz}. The parameter choices are listed in~\autoref{tab:H50_parameter_set}. Note that, in current work, the most relevant parameters are $m_5$ and $s_H$. The other parameters have minor influence. The choice of these parameters will be discussed in the following sections.

\begin{table}[!tbp]
\centering
\begin{tabular}{p{6cm} p{4cm}  p{4cm}}
\hline
SM Constants & Variable Parameters & Potential Parameters\\
\hline
$G_F = 1.1663787\times10^{-5}\,{\rm GeV}^{-2}$ & $m_5\in(10^{-3},10)\,\rm GeV$ & $\lambda_3=-10^{-5}$ \\
$m_h = 125\,\rm GeV$ & $s_H\in(10^{-6},10^{0})$ &  $\lambda_4 = 0.1$ \\
& & $\lambda_5=-4\lambda_2=-4$\\
& & $M_2 = -10^{-8}\,\rm GeV$\\
\hline
\end{tabular}
\caption{Model parameter settings for this work.}
\label{tab:H50_parameter_set}
\end{table}

\subsection{Theoretical Constraints}

Several theoretical constraints much be imposed on the parameters which ensure that the GM model is physically consistent and include perturbative unitarity, bounded from below (BFB) and global stability.
\begin{itemize}
\item \textbf{Perturbative Unitarity}:~Quartic couplings satisfy the perturbative unitarity of 2 $\rightarrow$ 2  scattering process~\cite{Aoki:2007ah}
\begin{align}
    | \mathcal{M} | < 8 \pi
\end{align}
We compute the coefficients of quartic coupling. Based on the charge $Q$ and hypercharge $Y$, the coupling coefficients have  been classified into several categories. We can get

\begin{align}
    &|12 \lambda_{1}+14 \lambda_{3}+22\lambda_{4} \pm \sqrt{(12 \lambda_{1}-14\lambda_{3}-22 \lambda_{4})^2+144 \lambda_{2}^{2}} | < 8 \pi \nonumber \\
    & |4\lambda_{1}-2\lambda_{3}+4\lambda_{4}\pm\sqrt{(4\lambda_{1}+2\lambda_{3}-4\lambda_{4})^{2}+4\lambda_{5}^{2}}| < 8 \pi \nonumber \\
    & |16 \lambda_{3} + 8 \lambda_{4}| < 8 \pi \nonumber \\
    & |4\lambda_{3} + 8 \lambda_{4} | < 8 \pi \nonumber \\
    & |4\lambda_{2}-\lambda_{5}| < 8 \pi \nonumber \\
    & |4 \lambda_{2} + 2 \lambda_{5}| < 8 \pi \nonumber \\
    & |4 \lambda_{2} -4 \lambda_{5}| < 8 \pi
\end{align}
\item \textbf{Bounded from Below}:~In order to maintain the stability of system, it is required that the potential is bounded from below. The constraint on parameter space is then given by~\cite{Arhrib:2011uy}
\begin{align}
    \lambda_1 &> 0, \nonumber \\
	\lambda_4 &> \left\{ \begin{array}{l l}
		- \frac{1}{3} \lambda_3 &  \ \lambda_3 \geq 0, \\
		- \lambda_3 &  \ \lambda_3 < 0, \end{array} \right. \nonumber \\
	\lambda_2 &> \left\{ \begin{array}{l l}
		\frac{1}{2} \lambda_5 - 2 \sqrt{\lambda_1 \left( \frac{1}{3} \lambda_3 + \lambda_4 \right)} &
			 \ \lambda_5 \geq 0 \ {\rm and} \ \lambda_3 \geq 0, \\
		\omega_+(\zeta) \lambda_5 - 2 \sqrt{\lambda_1 ( \zeta \lambda_3 + \lambda_4)} &
			 \ \lambda_5 \geq 0 \ {\rm and} \ \lambda_3 < 0, \\
		\omega_-(\zeta) \lambda_5 - 2 \sqrt{\lambda_1 (\zeta \lambda_3 + \lambda_4)} &
		 \ \lambda_5 < 0.
			\end{array} \right.
\end{align}
where
\begin{align}
\omega_{\pm}(\zeta) &= \frac{1}{6}(1 - B) \pm \frac{\sqrt{2}}{3} \left[ (1 - B) \left(\frac{1}{2} + B\right)\right]^{1/2},\\
\zeta &\in [\frac{1}{3},1],\\
	B &\equiv \sqrt{\frac{3}{2}\left(\zeta - \frac{1}{3}\right)} \in [0,1].
\end{align}

\item \textbf{Global stability}:~The global minimum of the scalar potential corresponds to the vacuum state with the lowest energy. Global stability requires that, for one thing, the global minimum of the scalar potential must be located at $\langle\phi\rangle\neq 0$ to ensure EWSB, and for another thing, the global minimum needs to maintain custodial $\rm{SU(2)}$ symmetry to ensure the consistency between theoretical analysis and experimental measurements of the electroweak parameter $\rho=1$. Apply the above two constraints to GM model parameters to ensure the stability of the global minimum (vacuum state).
\end{itemize}

\subsection{The Loop induced FCNC couplings for $H_5^0$}
\label{sec:FCNC_coefficient}

\begin{figure}[!h]
\centering
\includegraphics[width=0.95\textwidth]{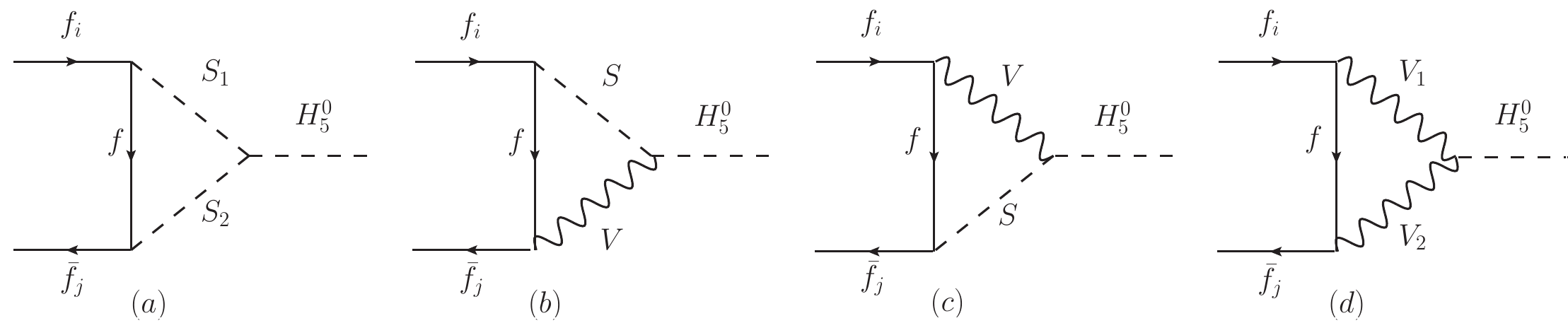}
\caption{The generic Feynman diagrams contributing to the FCNC coupling of two fermions and GM neutral fiveplet from renormalizable Lagrangian.}
\label{fig:FCNC_FeynmanDiagram}
\end{figure}

The loop induced FCNC coupling between two fermions and neutral fiveplet $H_{5}^{0}$ can be written as\footnote{There actually are other forms of the interaction between fermions and scalars. However, as we will apply the coupling in the case where the fermions in the coupling can be treated as on-shell. Hence all other forms can be naturally reduce to the form listed here, with the usage of the fermion EOM.}
\begin{align}
    \label{equ:L_FCNC}
    \mathcal{L}\supset \sum_{ij}\xi_{ij}^{5}\frac{m_{f_j}}{v} H_{5}^{0} \bar{f}_i P_R f_j + h.c.
\end{align}
where $\xi_{ij}^5$ is the corresponding couplings.
\autoref{fig:FCNC_FeynmanDiagram} shows the four kinds of 1-loop Feynman diagrams inducing the interactions between two fermions and $H_5^0$. We categorize the diagrams according to the particles running in the loop: (a) $FS_1S_2$, (b)/(c) $FSV$, (d) $FV_1V_2$. In fact, for the most general scalar interacting with two fermions, in addition to the four diagrams shown in~\autoref{fig:FCNC_FeynmanDiagram}, there are two extra cases $SF_1F_2$ and $VF_1F_2$ which depend on the coupling between the external scalar with the fermions in the loop. Nevertheless, the fiveplet in GM model is fermophobic at tree-level. We thus do not include these two contributions.

The generic results, obtained from {\tt FeynArts}~\cite{Hahn:2000kx} and {\tt FormCalc}~\cite{Hahn:1998yk}, for the diagrams in~\autoref{fig:FCNC_FeynmanDiagram} are given in the following:

\begin{align}
&\xi_{ji}^{5,a} = -\frac{v}{16\pi^2m_{f_i}}g_{H_5^0 S_1 S_2}\left\{g_{\bar{F}_jf S_2}^Rg_{\bar{f}F_{i}S_1}^Rm_fC_0-g_{\bar{F}_jfS_2}^Rg_{\bar{f}F_iS_1}^Lm_{F_i}C_1-g_{\bar{F}_jfS_2}^Lg_{\bar{f}F_iS_1}^Rm_{F_j}C_2\right\}\nonumber \\
&\qquad(m_{F_i}^2,m_5^2,m_{F_j}^2,m_f^2,m_{S_1}^2,m_{S_2}^2)
\end{align}
\begin{align}
&\xi_{ji}^{5,b} = \frac{v}{16\pi^2m_{f_i}}\left\{-g_{\bar{F}_jfV}^Lg_{\bar{f}F_iS}^Rg_{H_5^0SV}^SB_0(m_5^2,m_S^2,m_V^2)-m_f\left(g_{\bar{F}_jfV}^Lg_{H_5^0SV}^S(g_{\bar{f}F_iS}^Rm_f-g_{\bar{f}F_iS}^Lm_{F_i})\right.\right.\nonumber \\
&\qquad \left.+\left(g_{\bar{F}_jfV}^Lg_{\bar{f}F_iS}^Lm_{F_i}-g_{\bar{F}_jfV}^Rg_{\bar{f}F_iS}^Rm_{F_j}\right)g_{H_5^0SV}^{H_5^0}\right)C_0(m_{F_i}^2,m_5^2,m_{F_j}^2,m_f^2,m_S^2,m_V^2)\nonumber \\
&\qquad+m_{F_i}\left(g_{\bar{F}_jfV}^Lg_{H_5^0SV}^S\left(g_{\bar{f}F_iS}^Lm_f-g_{\bar{f}F_iS}^Rm_{F_i}\right)\right.\nonumber \\
&\qquad+ \left.\left(g_{\bar{F}_jfV}^Lg_{\bar{f}F_iS}^Rm_{F_i}-g_{\bar{F}_jfV}^Rg_{\bar{f}F_iS}^Lm_{F_j}\right)g_{H_5^0SV}^{H_5^0}\right)C_1(m_{F_i}^2,m_5^2,m_{F_j}^2,m_f^2,m_S^2,m_V^2)\nonumber \\
&\qquad-\left(m_{F_j}g_{\bar{F}_jfV}^R\left(g_{\bar{f}F_iS}^L(g_{H_5^0SV}^{H_5^0}-g_{H_5^0SV}^S)m_{F_i}-g_{\bar{f}F_iS}^Rg_{H_5^0SV}^Sm_f\right)\right.\nonumber \\
&\qquad\left.\left.+g_{\bar{F}_jfV}^Lg_{\bar{f}F_iS}^R\left(g_{H_5^0SV}^{H_5^0}(m_5^2-m_{F_i}^2)-g_{H_5^0SV}^S(m_5^2-m_{F_i}^2-m_{F_j}^2)\right)\right)C_2(m_{F_i}^2,m_5^2,m_{F_j}^2,m_f^2,m_S^2,m_V^2)\right\}
\end{align}
\begin{align}
&\xi_{ji}^{5,c} = -\frac{v}{16\pi^2m_{f_i}}\left\{-g_{\bar{F}_jfS}^Rg_{\bar{f}F_iV}^Rg_{H_5^0SV}^SB_0(m_5^2,m_S^2,m_V^2) - m_f\left(g_{\bar{f}F_iV}^Rg_{H_5^0SV}^S(g_{\bar{F}_jfS}^Rm_f-g_{\bar{F}_jfS}^Lm_{F_j})\right.\right.\nonumber \\
&\qquad \left.+ (g_{\bar{F}_jfS}^Lg_{\bar{f}F_iV}^Rm_{F_j}-g_{\bar{F}_jfS}^Rg_{\bar{f}F_iV}^Lm_{F_i})g_{H_5^0SV}^{H_5^0}\right)C_0(m_{F_j}^2,m_5^2,m_{F_i}^2,m_f^2,m_S^2,m_V^2)\nonumber \\
&\qquad + m_{F_j}\left(g_{\bar{f}F_iV}^Rg_{H_5^0SV}^S(g_{\bar{F}_jfS}^Lm_f-g_{\bar{F}_jfS}^Rm_{F_j})\right.\nonumber \\
&\qquad\left.+(g_{\bar{F}_jfS}^Rg_{\bar{f}F_iV}^Rm_{F_j}-g_{\bar{F}_jfS}^Lg_{\bar{f}F_iV}^Lm_{F_i})g_{H_5^0SV}^{H_5^0}\right)C_1(m_{F_j}^2,m_5^2,m_{F_i}^2,m_f^2,m_S^2,m_V^2)\nonumber \\
&\qquad -\left(m_{F_i}g_{\bar{f}F_iV}^L\left(g_{\bar{F}_jfS}^Lm_{F_j}(g_{H_5^0SV}^{H_5^0}-g_{H_5^0SV}^S)-g_{\bar{F}_jfS}^Rg_{H_5^0SV}^Sm_f\right) \right.\nonumber \\
&\qquad\left.\left.+g_{\bar{F}_jfS}^Rg_{\bar{f}F_iV}^R\left(g_{H_5^0SV}^{H_5^0}(m_5^2-m_{F_j}^2)-g_{H_5^0SV}^S(m_5^2-m_{F_j}^2-m_{F_i}^2)\right)\right)C_2(m_{F_j}^2,m_5^2,m_{F_i}^2,m_f^2,m_S^2,m_V^2)\right\}
\end{align}
\begin{align}
&\xi_{ji}^{5,d} = -\frac{v}{8\pi^2m_{f_i}}g_{H_5^0S_1S_2}\left\{2g_{\bar{F}_jfS_2}^Lg_{\bar{f}F_iS_1}^Rm_fC_0 +\ g_{\bar{F}_jfS_2}^Lg_{\bar{f}F_iS_1}^Lm_{F_i}C_1 + g_{\bar{F}_jfS_2}^Rg_{\bar{f}F_iS_1}^Rm_{F_j}C_2\right\}\nonumber \\
&\qquad(m_{F_i}^2,m_5^2,m_{F_j}^2,m_f^2,m_{S_1}^2,m_{S_2}^2)
\end{align}
where $m_X$ represents the mass of the corresponding particle $X$ in the diagram~\autoref{fig:FCNC_FeynmanDiagram}. $B_0, C_{0,1,2}$ are the loop functions following the {\tt LoopTools} conventions~\cite{Hahn:1998yk}. $g_{(\cdot)}$ represents the generic couplings between different particles given in the following conventions:
\begin{align}
    \mathcal{L} &\supset \bar{f}_i(g_{f_if_jS_k}^LP_L + g_{f_if_jS_k}^RP_R)f_jS_k + \frac{1}{p_{ijk}}g_{S_iS_jS_k}S_iS_jS_k + \bar{f}_i(g_{f_if_jV_k}^L\gamma_\mu P_L + g_{f_if_jV_k}^R\gamma_\mu P_R)f_jV_{k}^\mu \nonumber \\
    &\quad + (g_{S_iS_jV_k}^i k_{S_i}^\mu + g_{S_iS_jV_k}^j k_{S_j}^\mu)S_iS_jV_{k\,\mu} + \frac{1}{p_{ij}}g_{V_iV_jS_k}S_kV_i^\mu V_{j\,\mu}
\end{align}
where $p_{ijk}$ and $p_{ij}$ are the symmetric factor for the corresponding vertices, such that for all above vertices, $ig_{(\cdot)}$ will be the corresponding Feynman rule.

For $H_5^0$, which is the neutral component of the fiveplet in GM model, we list all the possible combinations of particles that can appear in the loops of~\autoref{fig:FCNC_FeynmanDiagram} in the calculations in~\autoref{tab:H50_particles}. Note that we are working in the Feynman gauge where all the goldstones ($G^\pm,G^0$) are included. All relevant couplings involved in the calculation are listed in~\autoref{app:couplings} according to their categories.

\begin{table}
\centering
\begin{tabular}{|c|c|c|}
\hline
Diagram & Particle Types & Particles \\
\hline\hline
\multirow[c]{6}{*}{(a)} & \multirow[c]{6}{*}{$S_1S_2f$}  & $G^0G^0f$ $^\triangle$  \\
& & $G^+G^-f$ \\
& & $G^+H_3^-f/H_3^+G^-f$ \\
& & $H_3^+H_3^-f$ \\
& & $G^0H_3^0f$ $^\triangle$/$H_3^0G^0f$  $^\triangle$\\
& & $H_3^0H_3^0f$  $^\triangle$\\
\hline
\multirow[c]{4}{*}{(b) and (c)} & \multirow[c]{4}{*}{$SVf$} & $G^0Zf$ $^\triangle$\\
& & $H_3^0Zf$  $^\triangle$\\
& & $G^\pm W^\mp f$ \\
& & $H_3^\pm W^\mp f$\\
\hline
\multirow[c]{2}{*}{(d)} & \multirow[c]{2}{*}{$V_1V_2f$} & $ZZf$  $^\triangle$ \\
& & $W^+W^-f$ \\
\hline
\end{tabular}
\caption{The particles in the loops for the diagrams with $H_5^0$. In general, $f$ indicates the fermion which can be inferred straightforwardly from the other particles involved in the diagram. Those indicated with $\triangle$ only appear when the two external fermions are of the same flavor.}
\label{tab:H50_particles}
\end{table}

 \section{The Productions and Decays of light CP-even fiveplet $H_5^0$ }
\label{Sec:Production_decay_H50}

\subsection{Effective Lagrangian}

In order to consider the production and decay of light CP-even fiveplet $H_5^0$, the interactions between $H_5^0$ and other SM particles are given as:
\begin{align}
    \mathcal{L}&\supset\xi^{5}_{W} \frac{2m_{W}^{2}}{v} H_{5}^{0} W^{\mu +}W^{-}_{\mu}+ \xi^{5}_{Z} \frac{m_{Z}^{2}}{v} H_{5}^{0} Z^{\mu} Z_{\mu}+\xi^{5}_{\gamma}\frac{\alpha_{ew}}{4\pi v} H_{5}^{0} F_{\mu\nu}F^{\mu\nu} + \xi_{ij}^5\frac{m_{f_j}}{v}H_5^0\bar{f}_iP_Rf_j + h.c.
\end{align}
where $\xi_{ij}^5$ has been given in the last section. The couplings between $H_5^0$ and gauge bosons are given as:
\begin{align}
\xi^{5}_{W}&=\frac{s_{H}}{\sqrt{3}}
\label{equ:GM_xi_W}\\
\xi^{5}_{Z}&=-\frac{2}{\sqrt{3}} s_{H}
\label{equ:GM_xi_Z}\\
\xi^{5}_{\gamma}
&= \xi^{5}_{W}\,\mathcal{A}_{1}^{\phi}(\tau_{W}^{5})
 - \sum_{j \in \{H_5^{\pm\pm}, H_5^\pm, H_3^\pm\}}\!
   Q_j^2\, \xi^5_{j}\, \frac{v}{2 m_j^2}\,
   \mathcal{A}_0^\phi(\tau_j^5)
\label{equ:GM_xi_gamma}
\end{align}

with $Q$ corresponds the scalar charge and $\tau_{i}^{5}=m_{5}^{2}/(4m_{i}^{2})$. The scalar interactions involved in $\xi^{5}_{\gamma}$ are
\begin{align}
\xi^5_{H_5^{\pm}}&=-\sqrt{3}vs_H\lambda_3+\sqrt{6}M_2 \\
\xi^5_{H_5^{\pm\pm}}&= 2\sqrt{3}vs_H\lambda_3-2 \sqrt{6}M_2\\
\xi^5_{H_3^{\pm}}&=-\frac{\sqrt{3}}{3}c_H^2s_H\lambda_3v + \frac{\sqrt{3}}{6}s_H(1+3c_H^2)\lambda_5v - \frac{\sqrt{6}}{6}s_H^2M_1 - \sqrt{6}c_H^2M_2
\end{align}
The form factors $\mathcal{A}_{0,1}^\phi$ are given as
\begin{align}
\mathcal{A}_{0}^{^\phi}(\tau)& =-\frac{1}{2}[\tau-f(\tau)]\tau^{-2}\\
    \mathcal{A}_{1}^{\phi}(\tau)&=-\frac{1}{2}[2\tau^{2}+3\tau+3(2\tau-1)f(\tau)]\tau^{-2}\\
    f(\tau)&=\begin{cases}
    \arcsin ^{2} \sqrt{\tau} &  \tau \leqslant 1 \\
    -\frac{1}{4}\left(\log \frac{1+\sqrt{1-1 / \tau}}{1-\sqrt{1-1 / \tau}}-i \pi\right)^{2} & \tau>1
    \end{cases}
\end{align}

\subsection{Productions of light CP-even fiveplet $H_{5}^{0}$ }
\label{sec:H5Production}
In our analysis, we focus on the forward region where the light scalars can be produced mainly from the decay of hadrons. The produced hadrons spectrum from $pp$ collision in the forward region are obtained from {\tt FORESEE}~\cite{Kling:2021fwx}. In the following, we will briefly introduce the most important channels for light scalar productions.

\subsubsection*{$B$ Meson Decays}

The decay of $B$ meson ($B^{0}$, $B^{\pm}$) into light CP-even fiveplet $H_{5}^{0}$ is primarily governed by $\xi_{bs}^5$. The ratio of decay rates is given by~\cite{Grinstein:1988yu,Chivukula:1988gp,Boiarska:2019jym}
\begin{align}
    \frac{{\rm Br}(B^{0,\pm} \rightarrow K^{0,\pm} H_{5}^{0})}{{\rm Br}(B^{0,\pm} \rightarrow D^{\pm,0}e^{\mp}\nu_{e})}&=\frac{12 \pi^{2} v^{2}}{m_{b}^{2}}\left(1-\frac{m_{5}^{2}}{m_{b}^{2}}\right)^2\frac{1}{f(m_{c}^{2}/m_{b}^{2})} \left|\frac{\xi^{5}_{bs}}{V_{cb}} \right| ^{2} ,
\end{align}
where $f(x)=(1-8x+x^2)(1-x^2)-12x^2\ln x$ is the phase space factor. ${\rm Br}(B^{0} \rightarrow D^{\pm}e^{\mp}\nu_{e})=0.101$ and ${\rm Br}(B^{\pm} \rightarrow D^{0}e^{\pm}\nu_{e})=0.108$ from Ref.~\cite{ParticleDataGroup:2022pth} are used.

\subsubsection*{Kaon Decays}

The decay of Kaon can also produce light CP-even fiveplet $H_5^0$ with the branching ratio given as~\cite{Feng:2017vli,Boiarska:2019jym,Gunion:1989we,Leutwyler:1989xj}
\begin{align}
    {\rm Br}(K^{\pm} \rightarrow \pi^{\pm} H_{5}^{0})&=\frac{2p_5^0}{\Gamma_{K^{\pm}}}\frac{|\mathcal{M}|^{2}}{16\pi m_{K^{\pm}}^2},\\
    {\rm Br}(K_{L} \rightarrow \pi^{0} H_{5}^{0})&=\frac{2p_5^0}{\Gamma_{K_{L}}}\frac{{\rm Re}(\mathcal{M})^{2}}{16\pi m_{K_{L}}^2},\\
    {\rm Br}(K_{S} \rightarrow \pi^{0} H_{5}^{0})&=\frac{2p_5^0}{\Gamma_{K_{S}}}\frac{{\rm Im}(\mathcal{M})^{2}}{16\pi m_{K_{S}}^2},
\end{align}
where $p_5^0=\lambda^{1/2}(m_{K}^{2},m_{\pi}^{2},m_{5}^{2})/(2m_{K})$ is the momentum of $H_{5}^{0}$ in the rest frame of the Kaons with $\lambda$ the K\"all\'en function:
\begin{align}
\lambda(a,b,c)=a^{2}+b^{2}+c^{2}-2(ab+ac+bc)
\label{equ:lambda_function}
\end{align}

The scattering amplitude of the Kaon decaying into $H_5^0$ is affected by not only flavor-changing effect between $d$ and $s$ quarks, but also the four quarks interaction intermediated by the $W$ boson. Combining the two effects, the scattering amplitude can be expressed as~\cite{Boiarska:2019jym,Bezrukov:2009yw,Feng:2017vli,Leutwyler:1989xj,Gunion:1989we}
\begin{align}
     \mathcal{M} &=G_{F}^{1/2} 2^{1/4} \xi^{5}_W \left[\frac{7\omega(m_{K}^2+m_{\pi}^2-m_{5}^2)}{18}-\frac{7A_{K}m_{K}^2}{9}\right]+\frac{\xi^{5}_{ds}}{2v}m_{s}\frac{m_{K}^{2}-m_{\pi}^2}{m_{s}-m_{d}}f_0^{K \pi}(q^2),
\end{align}
where $\omega \simeq 3.1 \times 10^{-7}$, and $A_{K} \approx 0$~\cite{Feng:2017vli,Boiarska:2019jym,Gunion:1989we,Leutwyler:1989xj}. The form factor is given by~\cite{Ball:2004ye}
\begin{align}
    f_0^{K \pi}(q^2)=\frac{F_{0}^{K \pi}}{1-q^{2} / (m_{fit}^{K})^{2}},
\end{align}
with $m_{fit}^{K} \rightarrow \infty$ and $F_{0}^{K \pi}\approx0.96$~\cite{Boiarska:2019jym,Ball:2004ye,Marciano:1996wy}.

\subsubsection*{$\eta/\eta'$ Decays}
The branching ratios of $\eta^{(\prime)}$ decaying into $H_{5}^{0}$ are given by~\cite{Batell:2018fqo}
\begin{align}
    {\rm Br}(\eta^{(\prime)} \rightarrow \pi H_{5}^{0})=\frac{2p_{H_{5}^{0}}^0}{\Gamma_{\eta^{(\prime)}}}\frac{|g_{H_{5}^{0}\eta^{(\prime)}\pi}|^2}{16\pi m_{\eta^{(\prime)}}^2},
\end{align}
with $p_5^0$ the $H_5^0$ momentum in the $\eta^{(\prime)}$ rest frame. The coupling of $\eta^{(\prime)}$ with $H_5^0$ is obtained from chiral perturbation theory and can be written as~\cite{Egana-Ugrinovic:2019wzj}
\begin{align}
g_{H_{5}\eta^{(\prime)}\pi}=-\frac{1}{v}\left[m_u \xi^{5}_u -m_d \xi^{5}_d+\frac{2}{9}\left(m_u-m_d\right)\sum_{q=c, b, t}\xi^{5}_q\right]c_{H_{5}^{0}\eta^{(')}\pi}\mathcal{B},
\label{equ:eta}
\end{align}
with $\mathcal{B}= m_{\pi}^{2}/(m_{u}+m_{d})$. $c_{H_{5}^{0}\eta^{(\prime)}\pi}$ depends on the $\eta-\eta'$ mixing angle $\theta_{\eta}\approx-13\degree$ as $c_{H_5^0\eta^{(\prime)}\pi} = (\cos\theta_{\eta} \pm \sqrt{2} \sin\theta_{\eta})/\sqrt{3}$~\cite{Batell:2018fqo,Domingo:2016yih}.

\subsubsection*{$\Upsilon$ Decays}
The $\Upsilon$ can also produce $H_5^0$ through radiative decay~\cite{Ellis:1985yb}, the branching fraction of $\Upsilon$ decaying to $\gamma$ and $H_{5}^{0}$ is given as~\cite{Winkler:2018qyg}
\begin{align}
{\rm Br}(\Upsilon\rightarrow\gamma H_{5}^{0})=\frac{ G_F m_b^2|{\xi^{5}_b}|^2}{\sqrt{2}\pi\alpha}\left(1-\frac{m_{5}^2}{m_\Upsilon^2}\right) \times \mathcal{F} \times {\rm Br}(\Upsilon\rightarrow e^+e^-),
\end{align}
where $\mathcal{F}= \frac{2}{3}  \left(1-m_{5}^6/m_\Upsilon^6\right)$ denotes the radiation correction function. ${\rm Br}(\Upsilon\rightarrow e^+e^-)\approx 0.0239$ is used~\cite{ParticleDataGroup:2022pth}.

\subsubsection*{Mesons Decay with Semileptonic Channels}
The $H_5^0$ can also be produced in the semileptonic decay channels of mesons.
The decay branching ratio in this case can be written as~\cite{Boiarska:2019jym,Chivukula:1988gp,Dawson:1989kr,Cheng:1989ib}
\begin{align}
    {\rm Br}(X \rightarrow H_{5}^{0} e\nu)=\frac{\sqrt{2}G_F m_{X}^{4} {|{\xi^{5}_W}|^2}}{96\pi^2 m_\mu^2(1-m_\mu^2/m_X^2)^2}f\left(\frac{m_{5}^2}{m_X^2}\right)\left(1-\frac{2n_h}{33-2n_l}\right)^2\times {\rm Br}(X\rightarrow\mu\nu),
\label{equ:semileptonic}
\end{align}
where $X=\pi^\pm, K^\pm, D^\pm, D_s^\pm$ and $f(x)$ is the form factor same as that in B meson decay. The number of heavy quarks $n_h$ and light quarks $n_l$ when describing different mesons are shown in~\autoref{tab:semileptonic_parameters} together with the corresponding leptonic decay branching ratio.

\begin{table}[!tbp]
    \centering
    \begin{tabular}{|c|c|c|c|c|}
        \hline
        $X$ & $\pi^{\pm}$ & $K^{\pm}$ & $D^{\pm}$ & $D_{s}^{\pm}$ \\  \hline\hline
        $n_h$ & 3 & 3 & 2 & 2 \\ \hline
        $n_l$ & 3 & 3 & 4 & 4 \\ \hline
        ${\rm Br}(X\rightarrow\mu\nu)$ & 0.999877 & 0.6356 & $3.74 \times 10^{-4}$ & $5.35 \times 10^{-3}$ \\ \hline
    \end{tabular}
    \caption{Relevant parameters for mesons decay with semileptonic channels}
    \label{tab:semileptonic_parameters}
\end{table}

\begin{figure}[!tbp]
        \centering
        \includegraphics[width=0.48\textwidth]{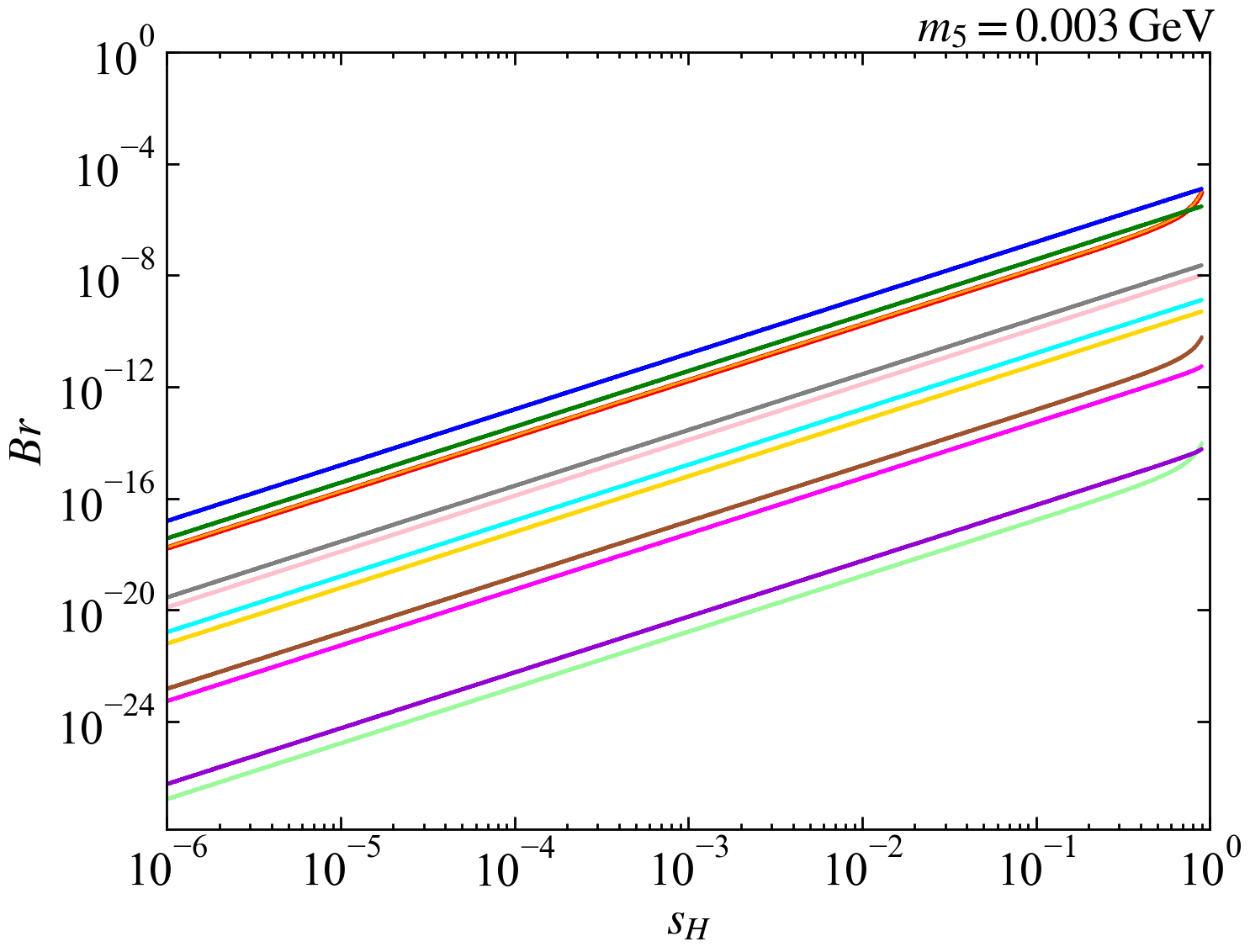}
        \includegraphics[width=0.48\textwidth]{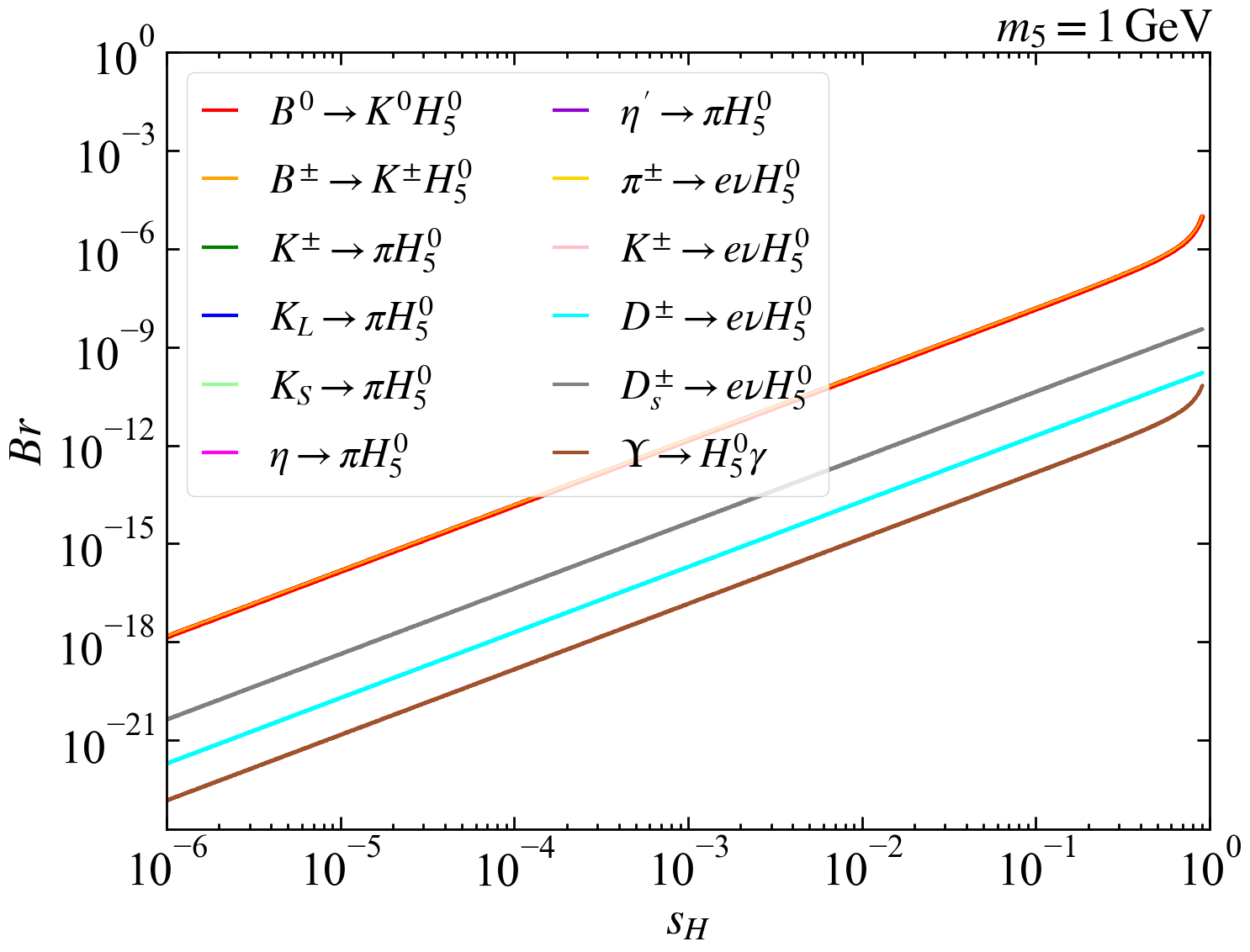}\\
        \includegraphics[width=0.48\textwidth]{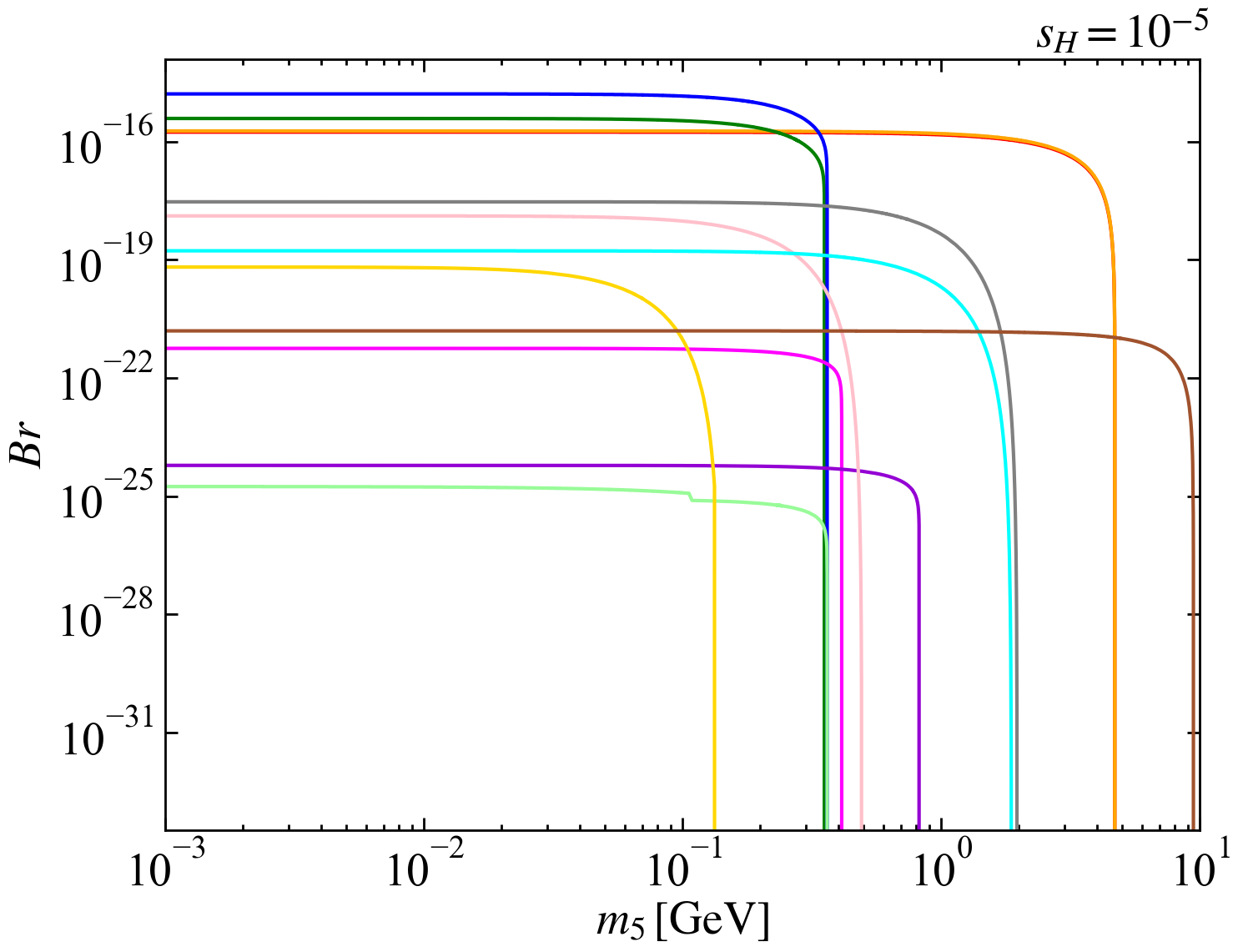}
        \includegraphics[width=0.48\textwidth]{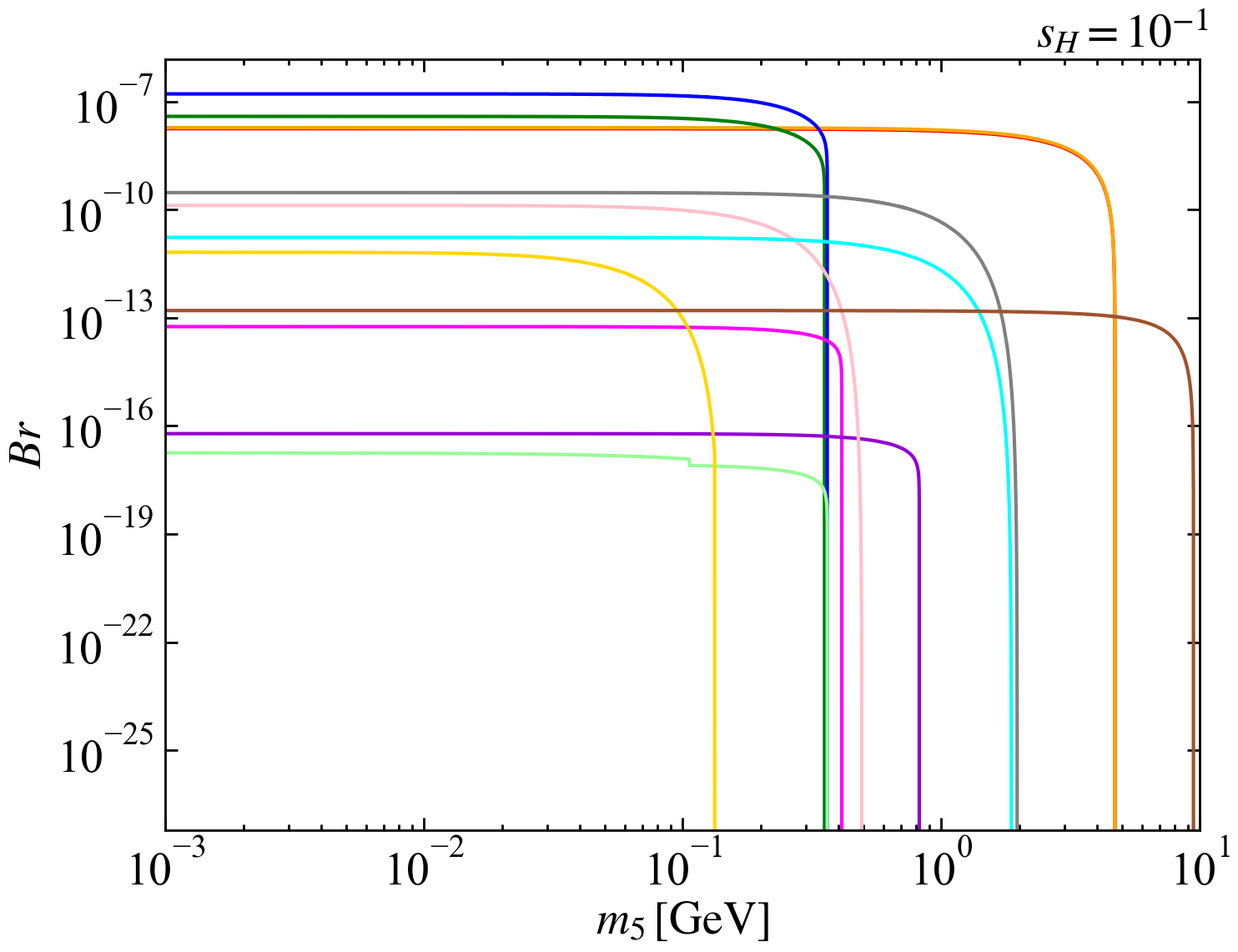}
        \caption{Decay branching ratios of different channels for the production of light CP-even fiveplet $H_5^0$ from meson decays as a function of $s_H$/$m_5$. The legend corresponding to the four figures is presented in the second figure. Notably, $Br(B^{0} \rightarrow K^{0} H_5^0)$ represented by the red line and $Br(B^{\pm} \rightarrow K^{\pm} H_5^0)$ represented by the orange line are found to be comparable, with the orange line slightly higher in value.}
	\label{fig:Br_meson_decay}
\end{figure}

\begin{figure}[!tbp]
\centering
\includegraphics[width=0.47\textwidth]{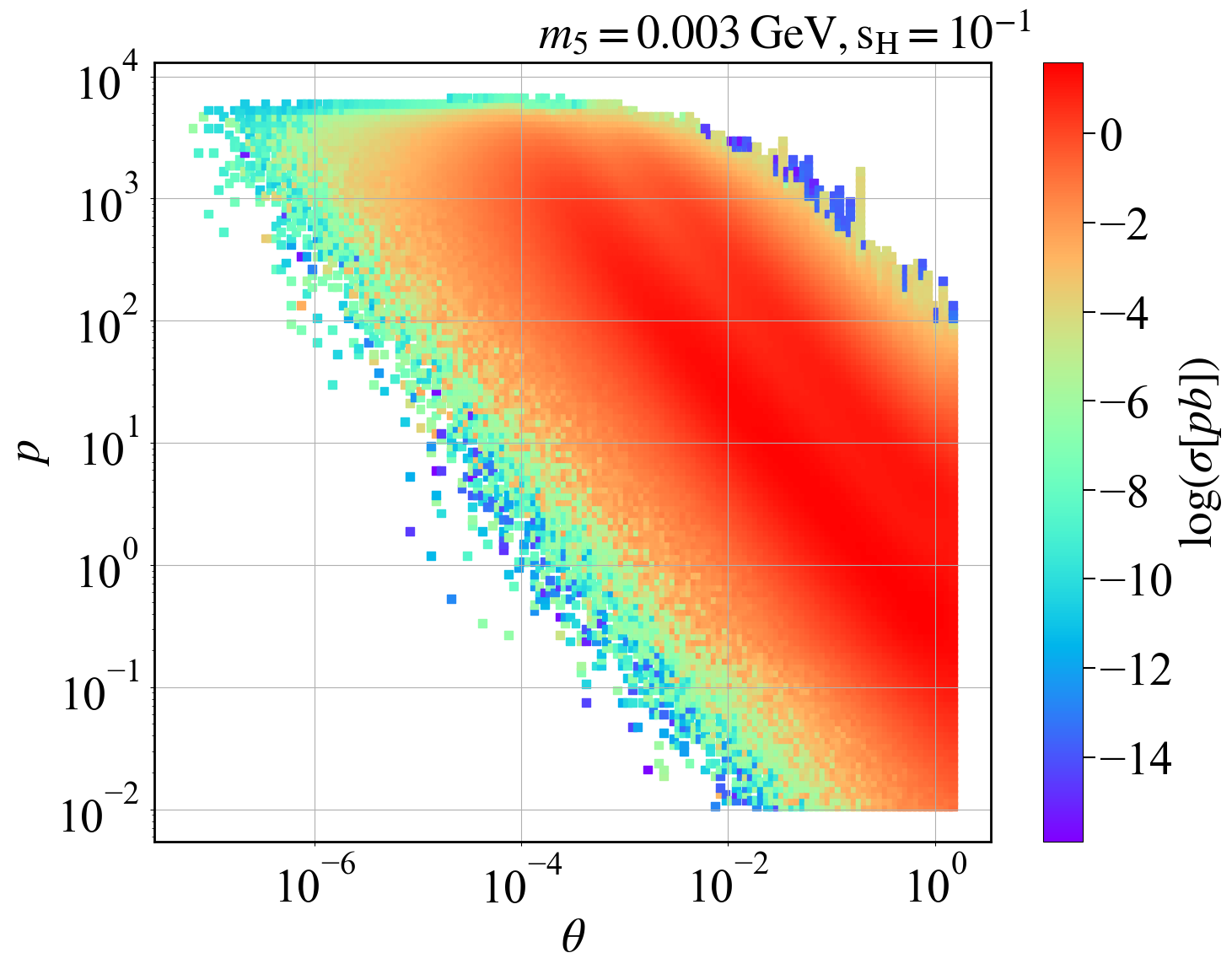}
\includegraphics[width=0.47\textwidth]{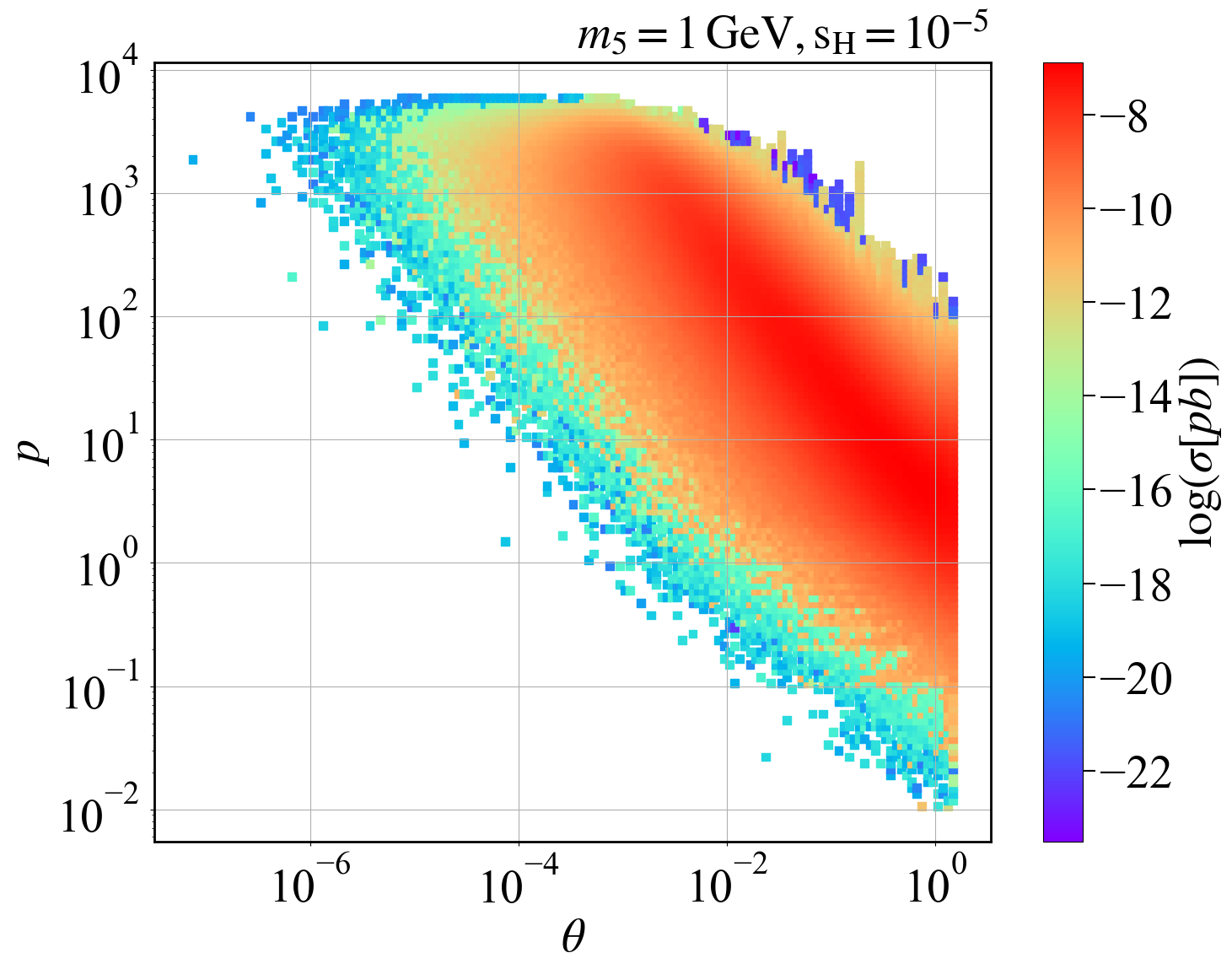}
\caption{The cross-section corresponding to the process of hadron collision producing mesons and mesons further decay into light CP-even fiveplet $H_5^0$.}
\label{fig:gm_scalar_rate}
\end{figure}

\begin{figure}[!tbp]
\centering
\includegraphics[width=0.6\textwidth]{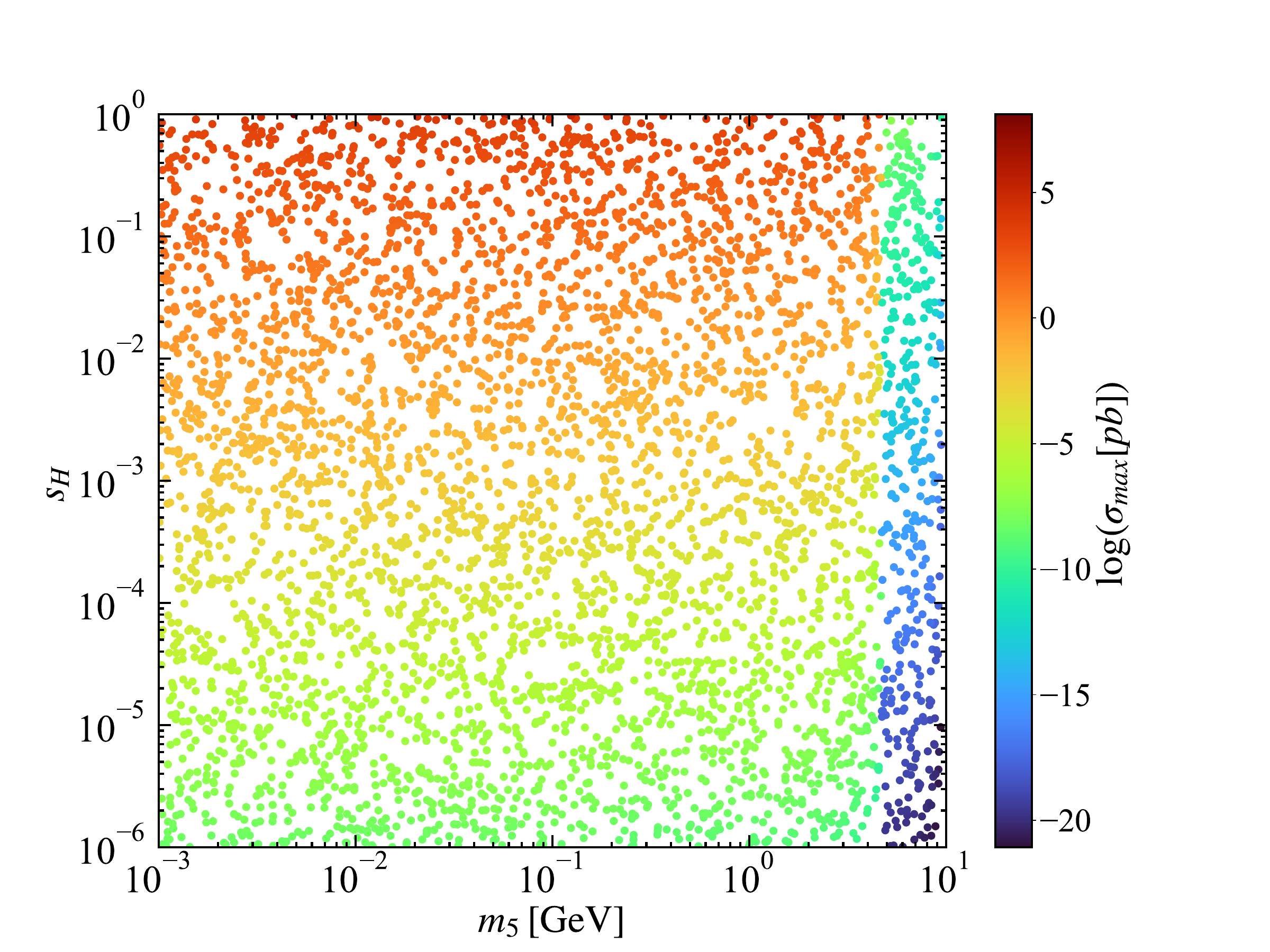}
\caption{The production cross-section of the light CP-even fiveplet $H_5^0$ in the $m_5$-$s_H$ parameter plane.}
\label{fig:sigma_m5_sH}
\end{figure}

The branching ratio of above meson decay channels for several benchmarks are shown in~\autoref{fig:Br_meson_decay} where in the upper panels $m_5$ is fixed and in the lower panel $s_H$ is fixed. Note that the other parameters have minor influence on the results. It is clear that all the branching ratios is proportional to $s_H^2$ (there will be tiny deviation when $s_H$ is large, where the influence of other parameters become large), while the dependence on $m_5$ is minor. However, when the mass is close to the threshold of particular decay channel, there will be a sudden change in the branching ratios. When the mass is small, the decay from K meson is dominant. In the heavier region, B mesons provide the dominant contribution. When the mass is heavier than several GeV, only $\Upsilon$ can produce $H_5^0$ through the radiative decay.
Combining with the production rate of the mesons from {\tt FORESEE}~\cite{Kling:2021fwx}, the production rate of $H_5^0$ for two benchmarks are shown in~\autoref{fig:gm_scalar_rate} in $\theta$-$p$ plane where $\theta$ is the polar angle and $p$ is the momentum of the scalar. The color represents the production cross section of $H_5^0$ in the specific $(\theta,p)$ bin. Although the branching ratios as shown in~\autoref{fig:Br_meson_decay} are very small, the production rate of the mesons are extremely large in the forward region,
the light scalar still has considerable production rate in the forward region. The maximum cross section of $H_5^0$ production in the binning given in~\autoref{fig:gm_scalar_rate} as function of $m_5$ and $s_H$ are shown in~\autoref{fig:sigma_m5_sH}. It is clear that the dependence on $m_5$ is minor until reaching the threshold, while the cross section drops dramatically with $s_H$.

\subsection{Decays of $H_{5}^{0}$ }
In the mass region we considered, $H_5^0$ mainly decays into photon pair, leptons, light hadrons and light quarks which will be briefly discussed in the following.

\subsubsection*{Decays into Pair of Photons}
The decay width of $H_{5}^{0}\to\gamma\gamma$ is given by
\begin{align}
    \Gamma(H_{5}^{0} \rightarrow \gamma \gamma)=\frac{G_{F} \alpha^{2} m_{5}^{3}}{32 \sqrt{2} \pi^{3}}  \left| \xi^{5}_\gamma \right|^{2},
\end{align}
where $\xi^{5}_\gamma$ is given in~\autoref{equ:GM_xi_gamma}. In our studies, the scalars other than the fiveplet are set to be heavy, which is achieved with large $\lambda_5$ as given in~\autoref{tab:H50_parameter_set}. Hence the triplet contribution to $\xi_\gamma^5$ is negligible. Under this condition, we have
\begin{align}
\label{equ:xi_5_gamma}
    \xi_\gamma^5 \approx \frac{s_H}{\sqrt{3}}\mathcal{A}_1^\phi(\tau_W^5) - 7\sqrt{3}(vs_H\lambda_3 -\sqrt{2}M_2) \frac{v}{2m_5^2}\mathcal{A}_0^\phi(\tau^5_{H_5})
\end{align}

\subsubsection*{Decays into Pair of Leptons}
Electron, muon and tau are all included whenever the phase space is allowed. The decay width for the process of $H_{5}^{0}$ decaying to leptons reads~\cite{Winkler:2018qyg}
\begin{align}
    \Gamma(H_{5}^{0} \rightarrow \ell^+ \ell^-) =\frac{G_F \beta_\ell^3 m_{5} m_\ell^2 }{4\sqrt{2}\pi} \left|\xi^{5}_{\ell} \right|^2.
\end{align}
with $\beta_\ell$ is $\sqrt{1-4m_\ell^2/m_{5}^2}$.

\subsubsection*{Decays into Mesons via Hadronic Processes}
If $m_{5} \lesssim 2 ~\rm{GeV}$~\cite{Donoghue:1990xh}, pions and kaons can be produced through hadronic decays. The decay width for the process is
\begin{align}
   \Gamma(H_{5}^{0}\rightarrow \pi \pi)&=\frac{3 G_{F}}{16 \sqrt{2} \pi m_{5}} \beta_{\pi}\left|
\frac{m_u \xi^{5}_{u}+m_d \xi^{5}_{d}}{m_u+m_d} \Gamma_{\pi} + (\xi^{5}_{s})\Delta_{\pi}
\right|^{2}, \\
\Gamma(H_{5}^{0}\rightarrow K K)&=\frac{ G_{F}}{4 \sqrt{2} \pi m_5} \beta_{K}\left|
\frac{m_u \xi^{5}_{u}+m_d \xi^{5}_{d}}{m_u+m_d} \Gamma_{K} + (\xi^{5}_{s})\Delta_{K}
\right|^{2},
\end{align}
where $\beta_{i}=\sqrt{1-4m_{i}^2/m_{5}^2}$ and the values of form factor $\Theta_{i},\Gamma_{i},\Delta_{i}$ ($i=\pi,K$) refer to~\cite{Winkler:2018qyg}.

\begin{figure}[!tbp]
		\centering
        \includegraphics[width=0.48\textwidth]{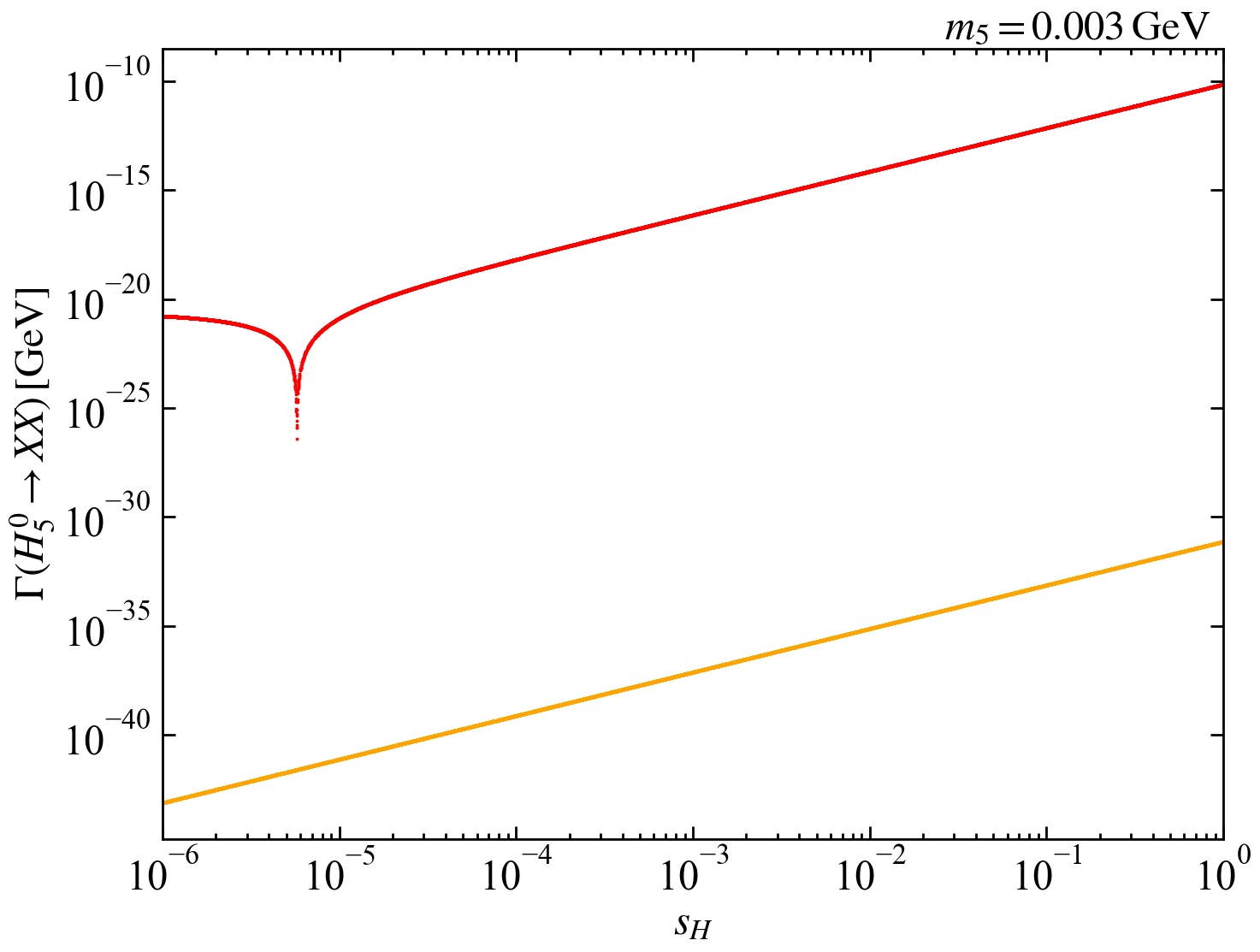}
        \includegraphics[width=0.48\textwidth]{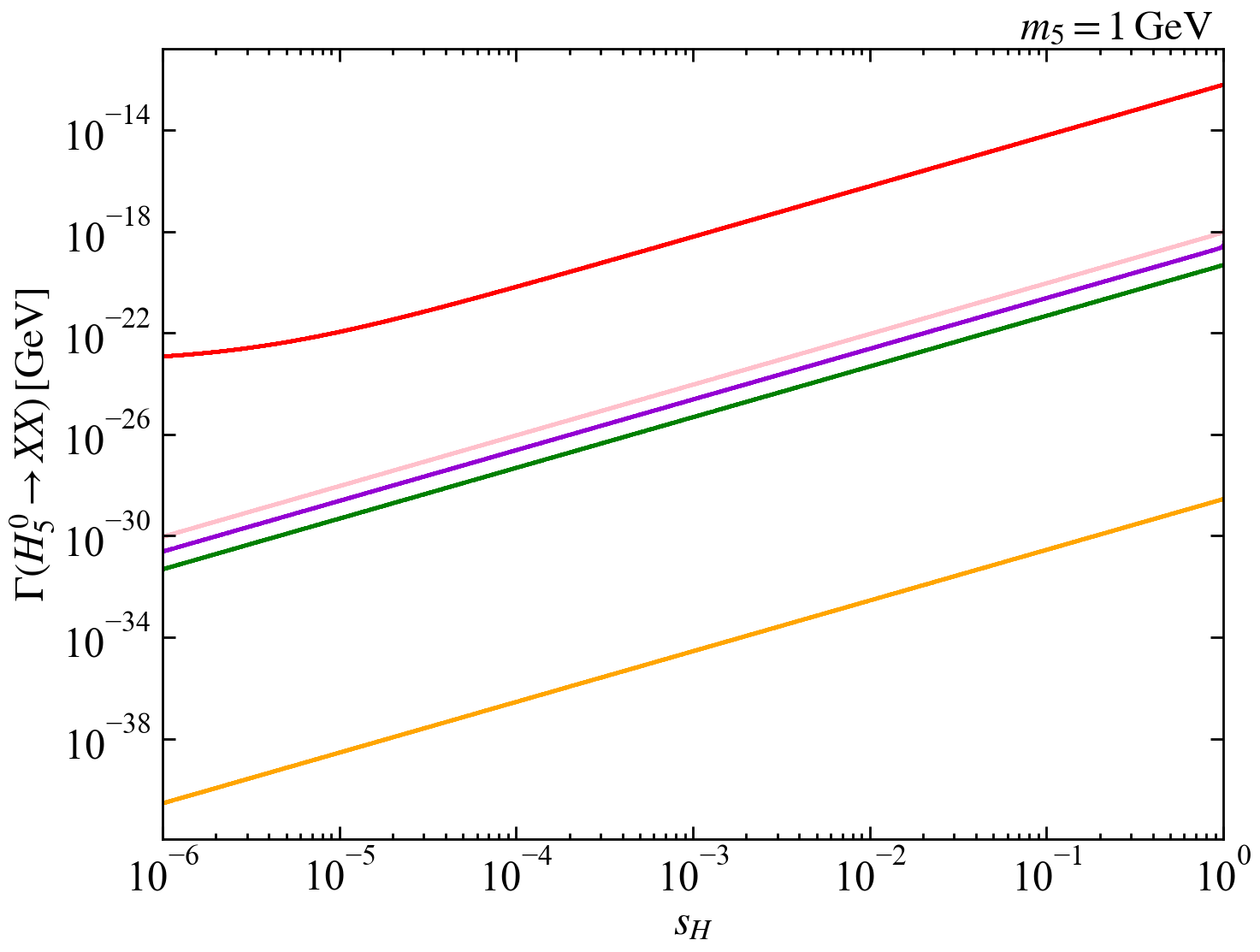}\\
        \includegraphics[width=0.48\textwidth]{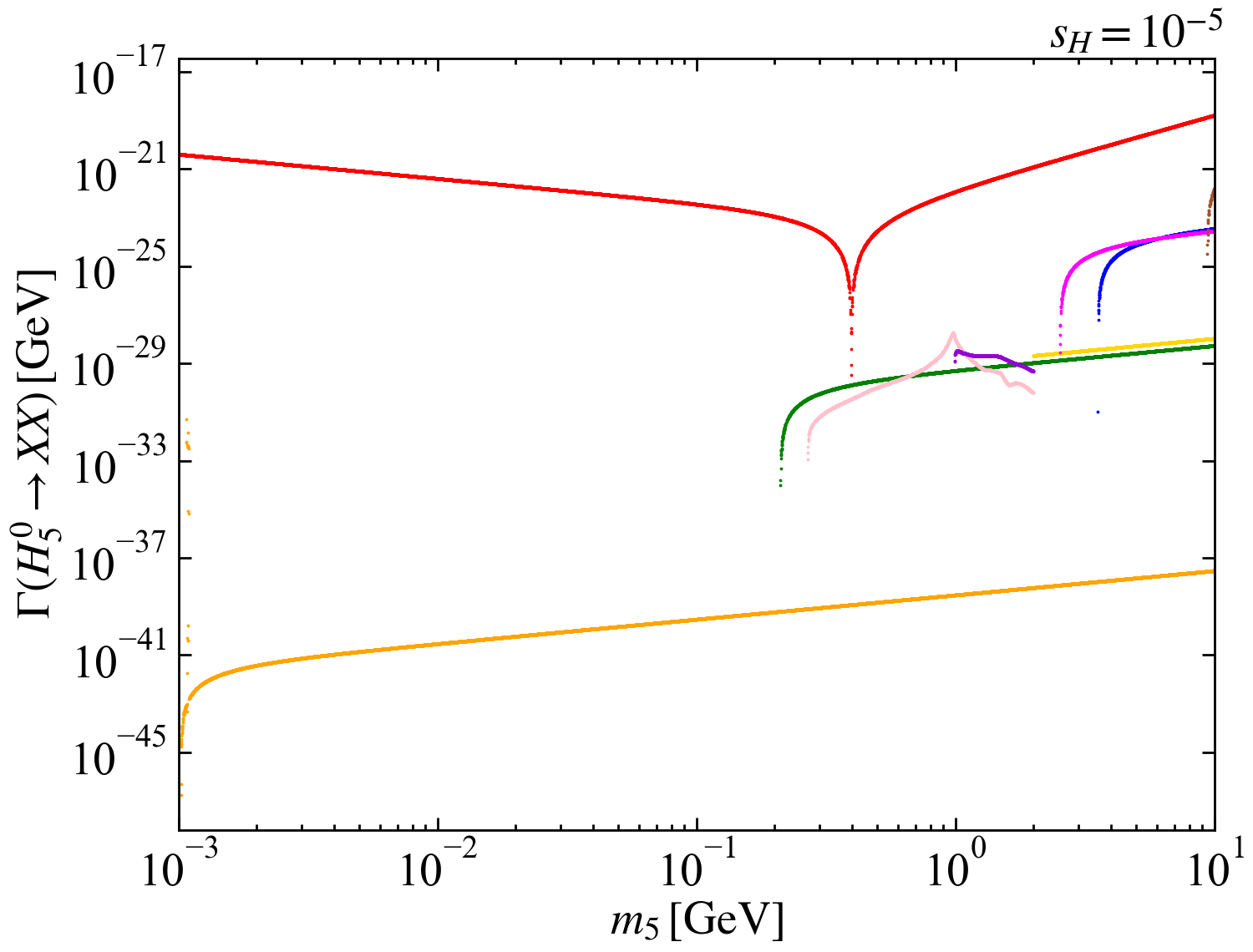}
        \includegraphics[width=0.48\textwidth]{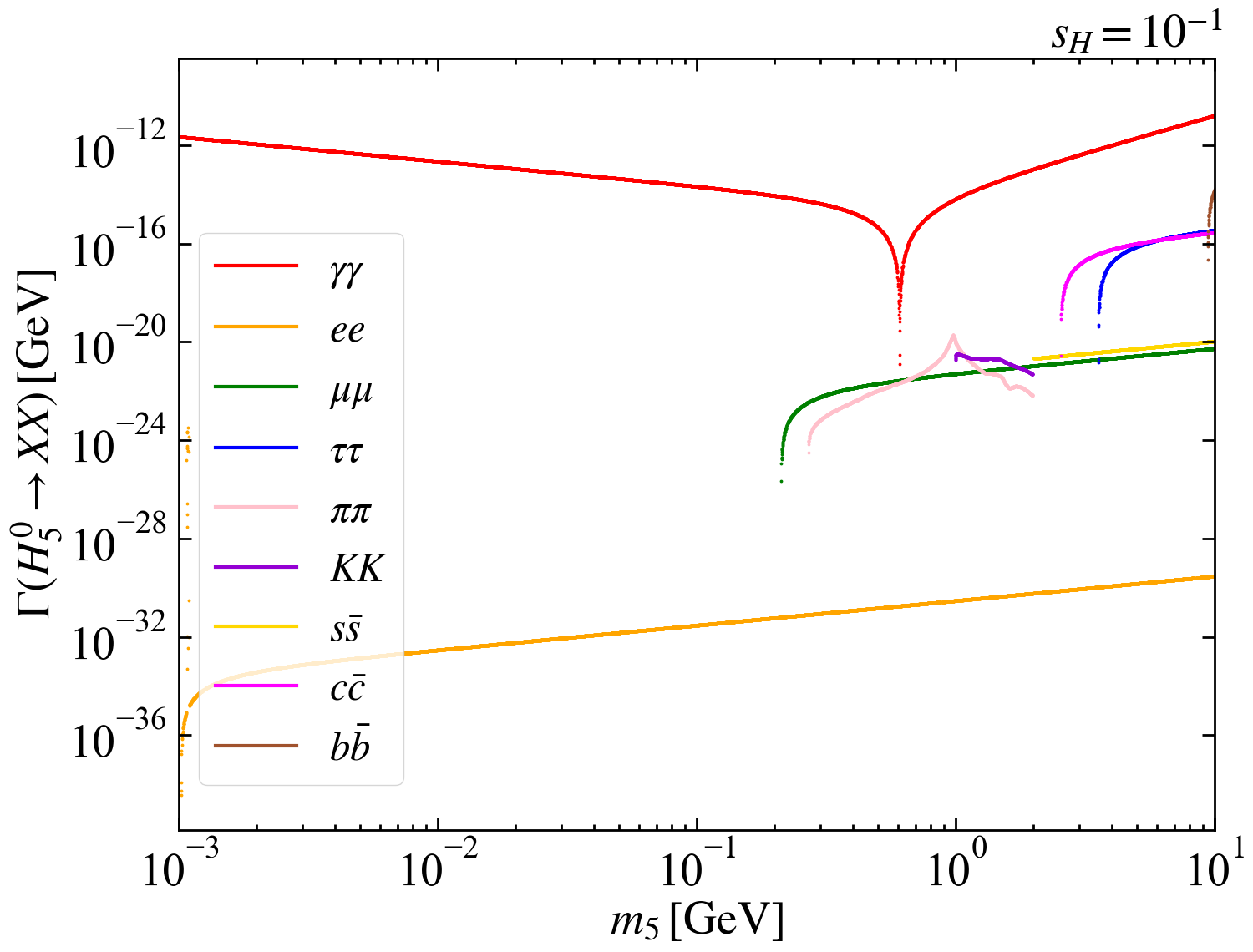}
     \caption{Decay widths of the light CP-even fiveplet $H_5^0$ into various final states as a function of $s_H$/$m_5$}
     \label{fig:H50_gamma}
\end{figure}

\subsubsection*{Decays into Pair of Quarks}
If $m_{5}\gtrsim 2 ~\rm{GeV}$, we can utilize the perturbative spectator model to calculate the decay width related to $H_{5}^{0}$ decaying into quarks~\cite{Winkler:2018qyg}.
\begin{align}
    \Gamma(H_5^0 \rightarrow s\bar{s})&=\frac{3|\xi^{5}_{s}|^2 m_s^2\beta_K^3}{|\xi^{5}_{\ell}|^2  m_\ell^2\beta_\ell^3}\Gamma_{\ell^+\ell^-},\\
    \Gamma(H_5^0 \rightarrow c\bar{c})&=\frac{3 |\xi^{5}_c|^2 m_c^2\beta_D^3}{|\xi^{5}_{\ell}|^2  m_\ell^2\beta_\ell^3}\Gamma_{\ell^+\ell^-},\\
    \Gamma(H_5^0 \rightarrow b\bar{b})&=\frac{3 |\xi^{5}_b|^2 m_b^2\beta_B^3}{|\xi^{5}_{\ell}|^2  m_\ell^2\beta_\ell^3}\Gamma_{\ell^+\ell^-},
\end{align}
where $\beta_{i}=\sqrt{1-4m_{i}^2/m_{5}^2}$ with $i=K, D, B$.

\subsubsection*{The Influence of the GM parameters}

All the parameters will affect all above decay channels, however, the dominant effects come from $m_5$ and $s_H$. We fix the parameters of the scalar potential according to~\autoref{tab:H50_parameter_set} and set $m_5$ and $s_H $ as follows:
\begin{itemize}
    \item Fixed $m_5=0.003/1\:\rm{GeV}$ with $s_H\in(10^{-6},10^{0})$
    \item Fixed $s_H=10^{-5}/10^{-1}$ with $m_5\in(10^{-3},10)\:\rm{GeV}$.
\end{itemize}
Based on the above setting, we calculate the deacy width of the light CP-even fiveplet $H_5^0$ which is shown in~\autoref{fig:H50_gamma}. In almost all the parameter space $H_5^0\to\gamma\gamma$ plays the leading role, hence, the total decay width of $H_5^0$ and consequently the decay length is mainly determined by the decay width of the diphoton channel. The exception exists in the region where different contributions in diphoton channel cancel with each other. In our case, there are two main contributions to $H_5^0\to\gamma\gamma$, one comes from the $W$-boson which is proportional to $s_H$, and the other comes from the fiveplets and is proportional to $(vs_H\lambda_3-\sqrt{2}M_2)$. Hence, the cancellation heavily depends on this parameter ($vs_H\lambda_3-\sqrt{2}M_2$) besides $m_5$ and $s_H$ and only happens when $vs_H\lambda_3-\sqrt{2}M_2<0$ for light $H_5$.

\begin{figure}[!btp]
		\centering
        \includegraphics[width=0.49\textwidth]{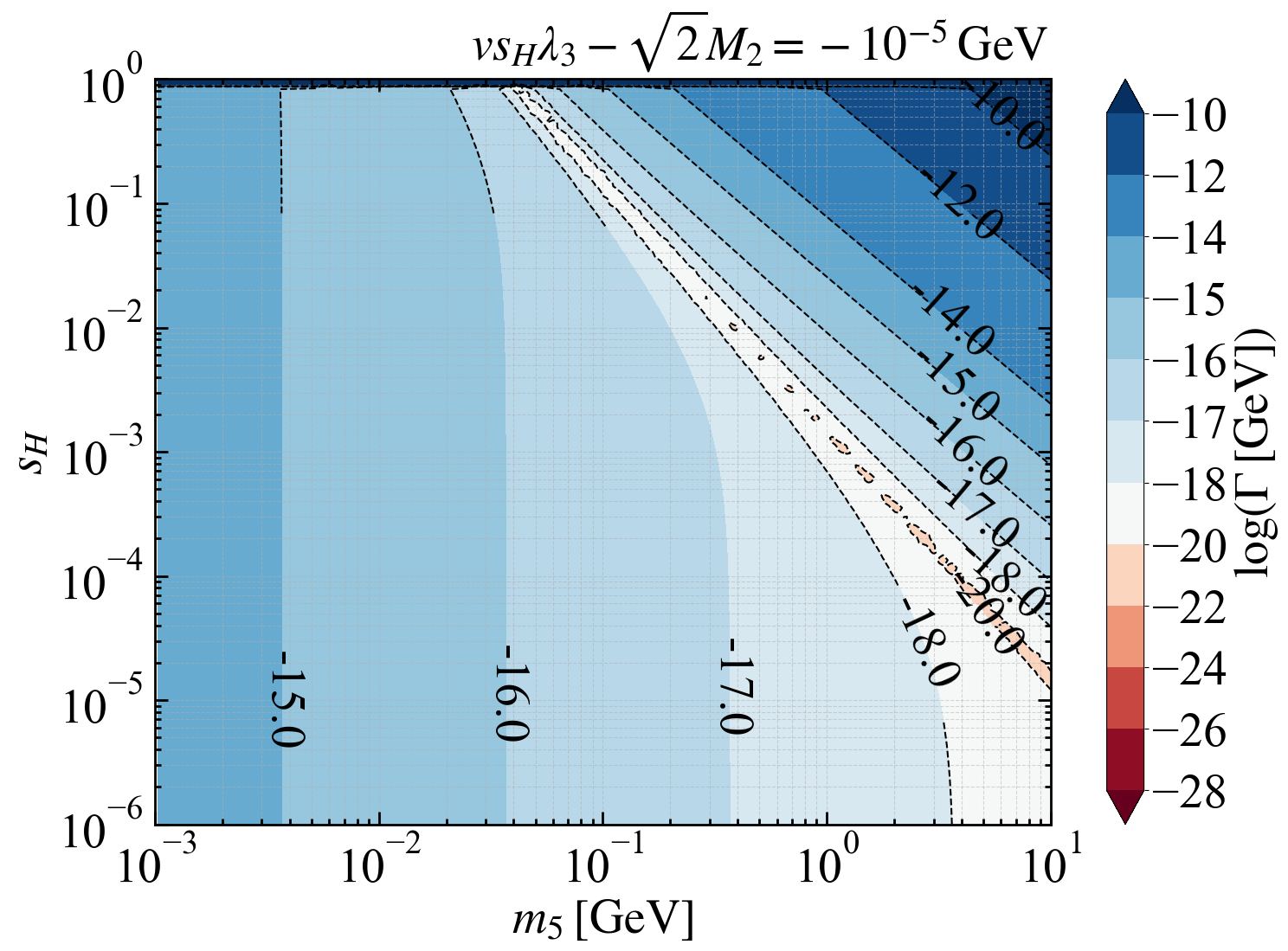}
        \includegraphics[width=0.49\textwidth]{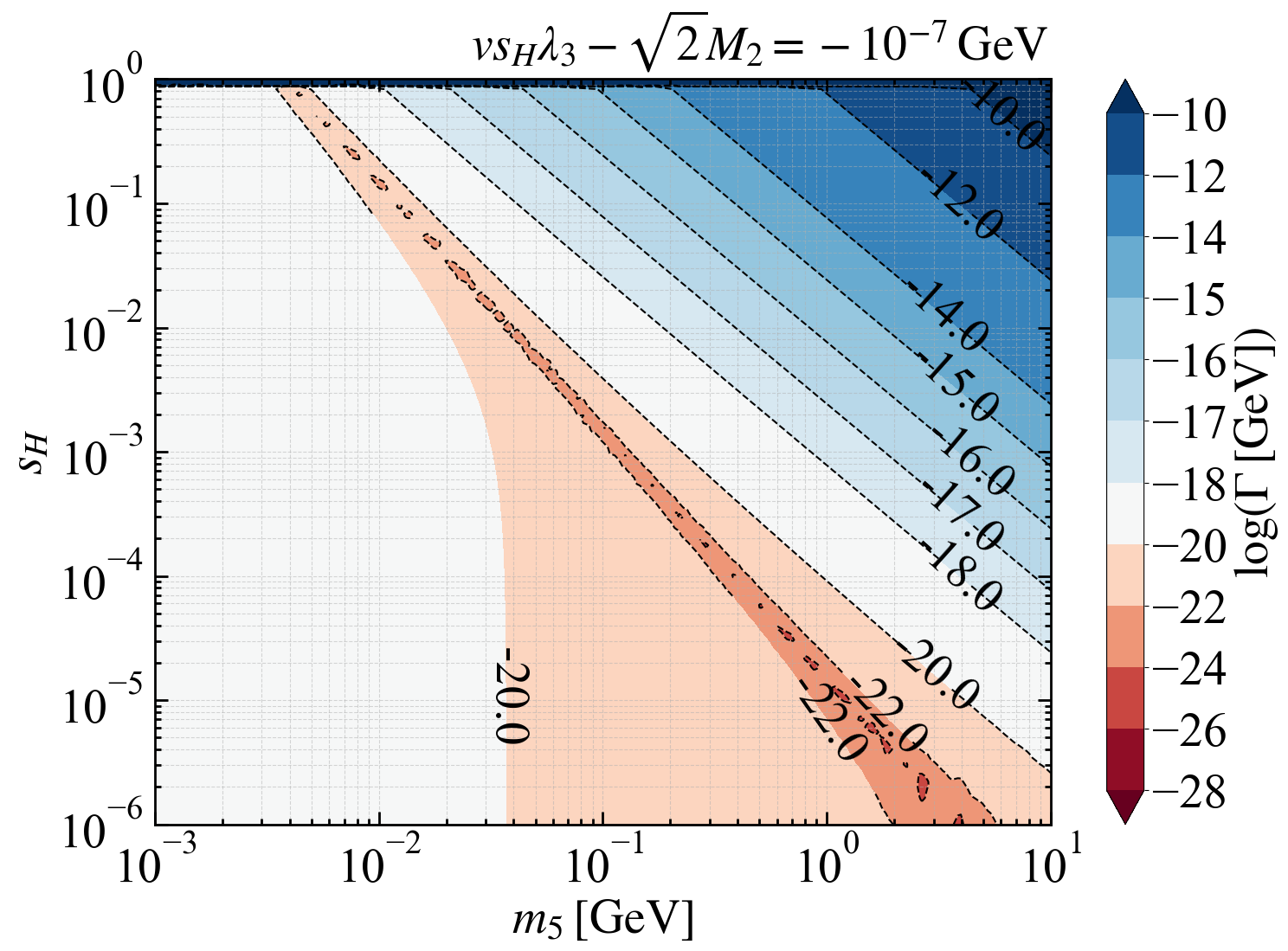}\\
        \includegraphics[width=0.49\textwidth]{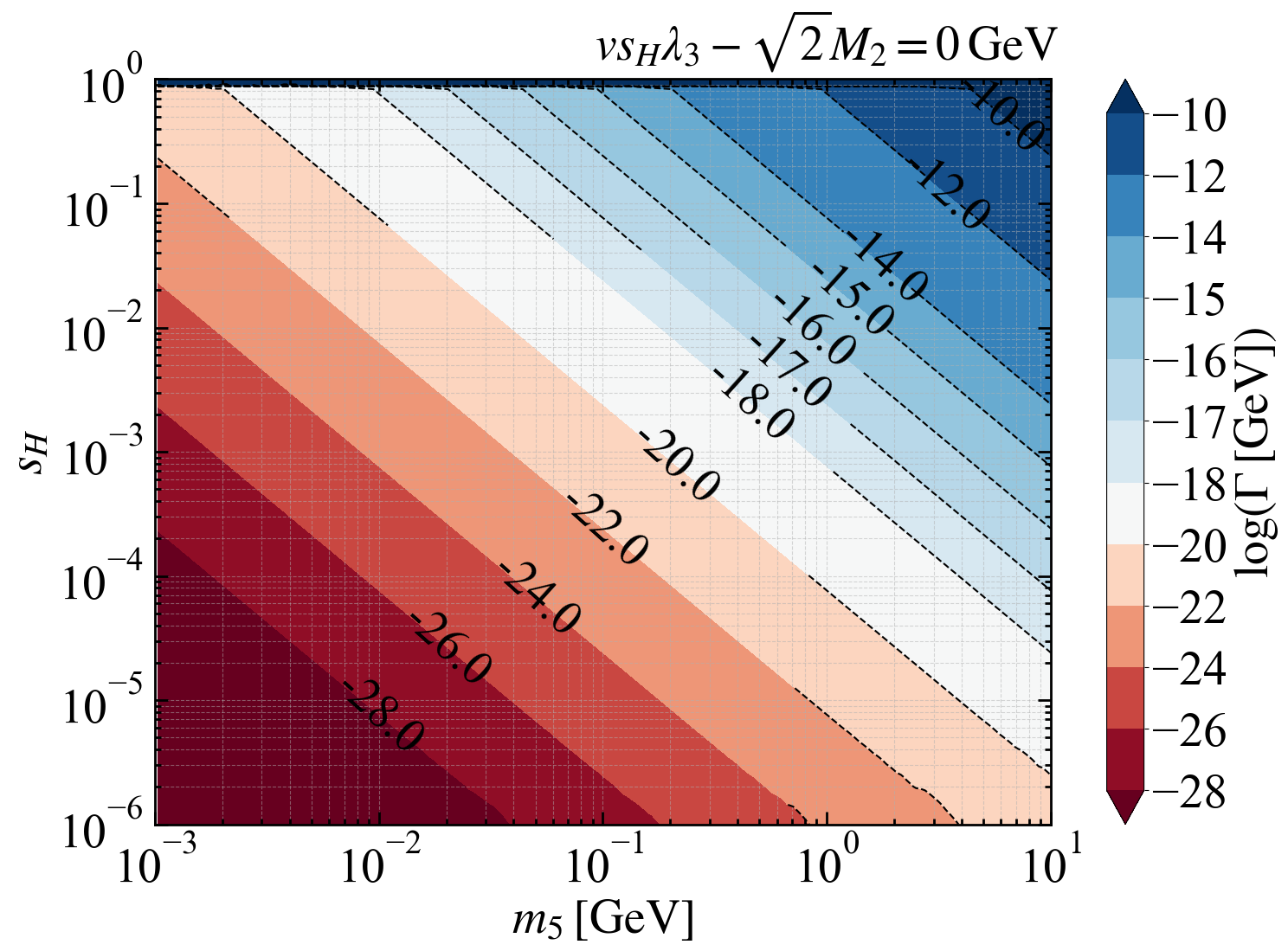}
        \includegraphics[width=0.49\textwidth]{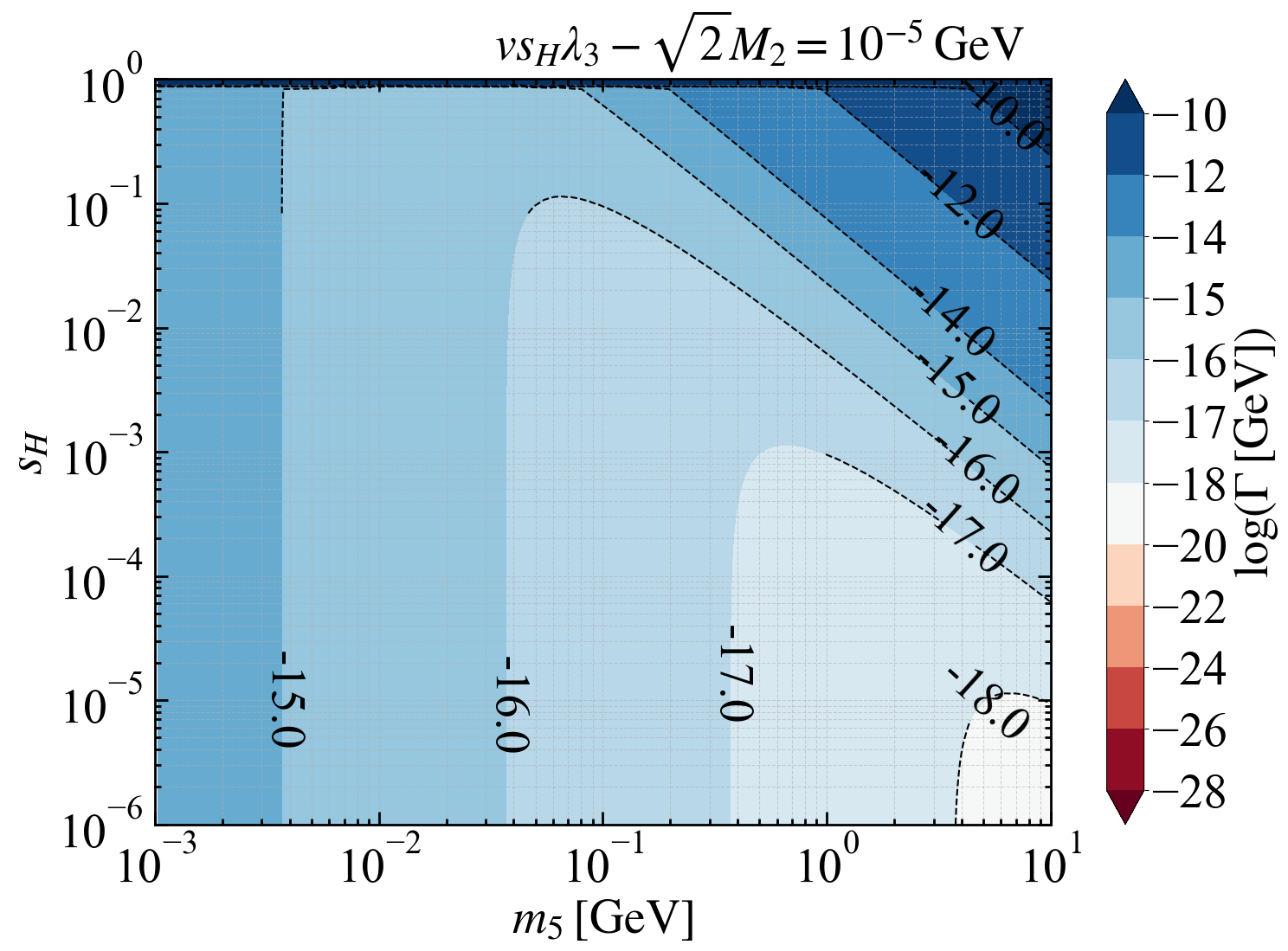}
     \caption{Decay widths of the light CP-even fiveplet $H_5^0$ with four GM model parameters.}
     \label{fig:H50_Gamma_vshlam3_M2}
\end{figure}

In order to investigate the detailed behavior of the total width (mainly from $H_5^0\to\gamma\gamma$), we choose four benchmarks for $vs_H\lambda_3-\sqrt{2}M_2$: $-10^{-5}$ GeV, $-10^{-7}$ GeV, $0$ GeV and $10^{-5}$ GeV. The total width of $H_5^0$ in $m_5$-$s_H$ plane for these four benchmarks are shown by the contours in~\autoref{fig:H50_Gamma_vshlam3_M2}. In the upper panels of~\autoref{fig:H50_Gamma_vshlam3_M2} where $vs_H\lambda_3-\sqrt{2}M_2<0$, we can clearly see the places where the cancellation happens. According to~\autoref{equ:xi_5_gamma}, we can roughly determine the place of cancellation for $vs_H\lambda_3-\sqrt{2}M_2<0$ as
\begin{align}
    \label{equ:cancellation}
    s_H \approx \frac{21v|vs_H\lambda_3-\sqrt{2}M_2|}{2m_5^2} \left|\frac{\mathcal{A}_0^5(\tau^5_{H_5})}{\mathcal{A}_1^5(\tau_W^5)}\right|.
\end{align}
On the other hand, the above equation also determines the boundary between the regions where either $W$ contribution or the scalar contribution in~\autoref{equ:xi_5_gamma} dominates. Specifically, when $s_H$ is smaller, the scalar contribution dominates. In this case, the total width is totally determined by $m_5$ when $vs_H\lambda_3-\sqrt{2}M_2$ is fixed as can be seen from~\autoref{fig:H50_Gamma_vshlam3_M2} where the contours in the panels for $vs_H\lambda_3-\sqrt{2}M_2=-10^{-5},\,-10^{-7},\,10^{-5}\,\rm GeV$ are vertical when the $s_H$ is below the value determined by~\autoref{equ:cancellation}. Note that for $vs_H\lambda_3-\sqrt{2}M_2=0$, we only have $W$ contribution which is also dominant in other panels where the contours are inclined.

\section{Searches for light CP-even fiveplet $H_{5}^{0}$}
\label{Sec:H50_analysis}

\subsection{Current Experimental Constraints}

\subsubsection*{Higgs Invisible Decay}
When the fiveplet is light, we will have the decay $h\to H_5H_5$, where $h$ is SM-like 125 GeV Higgs.
The current strictest limit on Higgs invisible decay is given as~\cite{Belanger:2013xza,CMS:2016dhk,ATLAS:2015gvj,ATLAS:2017nyv,ATLAS:2023tkt,Kling:2022uzy,Qi:2025qsj}
\begin{align}
{\rm Br}(h \rightarrow \text{invisible})<0.107~(\rm{at}~95\% ~\rm{C.L.})
\end{align}
In GM model, the decay branching ratio of $h \rightarrow H_{5}H_{5}$~\cite{Feng:2017vli}
\begin{align}
Br(h \rightarrow H_{5}H_{5})&=\frac{\Gamma(h \rightarrow H_{5}H_{5})}{\Gamma_{h}}=\frac{|g_{hH_{5}H_{5}}|^{2}}{8 \pi m_h \Gamma_{h}}\sqrt{1-\frac{4m_5^2}{m_h^2}}
\end{align}
where $\Gamma_h$ is the total width of the Higgs including all possible non-SM channels. The corresponding coupling of $hH_5H_5$ is given as
\begin{align}
g_{hH_{5}H_{5}}=i( 2\sqrt{3} M_{2} s_\alpha+8 \sqrt{3} (\lambda_{3}+\lambda_4) s_\alpha v_{\chi} -(4\lambda_2+\lambda_5)c_\alpha v_{\phi} ).
\end{align}
Note that the first two terms are suppressed by $s_\alpha$ and/or $v_\chi$, while the last term is not suppressed and can contribute significantly to the decay width of $h\to H_5H_5$. Hence, in our parameter choice listed in~\autoref{tab:H50_parameter_set}, we have $4\lambda_2+\lambda_5=0$ in order to have small Higgs invisible decays suppressed by $s_\alpha$ and $v_\chi$ (hence $s_H$).

\subsubsection*{Supernova}
The light CP-even fiveplet $H_5^0$ is also subject to constraints from supernova(SN) observations. During SN explosion, a light and weakly coupled scalar may be produced via the nucleon bremsstrahlung process $NN \to NN \phi$, where $N = n, p$ represents either a neutron or proton. If the scalar is weakly coupled, it can easily escape from the SN core carrying significant energy with sufficient flux which potentially changes the evolution of the SN~\cite{Turner:1988}. The observed neutrino signal from SN1987a core collapse indicates an upper bound on the possible energy loss rate as $\mathcal{P}\lesssim 10^{53}\,{\rm erg/s}$ which consequently provides indirect limits on the couplings between the scalar and the nucleons~\cite{ELLIS1987525, Krnjaic:2015mbs,Turner:1988, Frieman:1987, Burrows:1989, Essig_2010}.

The energy loss rate per unit volume due to scalar $\phi$ production is given by~\cite{Ishizuka:1990,Krnjaic:2015mbs}:
\begin{align}
Q_\phi &\sim C_{\phi N}^{2} \frac{11}{(15\pi)^{3}} \left(\frac{T}{m_{\pi}}\right)^{4} p_{F}^{5} G_{\phi} \left(\frac{m_{\pi}}{p_{F}}\right) \xi(T,\boldsymbol{m}_{\phi}),
\end{align}
where $T \sim 30\,\mathrm{MeV}$ is the temperature of the SN, $m_{\pi}$ is the pion mass, and $p_F \sim 200\,\mathrm{MeV}$ is the Fermi momentum of the SN. The effective coupling with nucleons can be written as~\cite{SHIFMAN1978443}:
\begin{align}
C_{\phi N} &= \frac{m_{N}}{v} \left( \sum_{q=u, d, s} \text{Re}(\xi_{q}^{\phi}) f_{Tq}^{(N)} + \frac{2}{27} f_{TG}^{(N)} \sum_{q=c, b, t} \text{Re}(\xi_{q}^{\phi}) \right),
\end{align}
where $\xi^\phi_{q} $ is the coupling between $\phi$ and SM quark $q$. Additionally, $f_{TG}^{(N)}, f_{Tq}^{(N)} \sim \mathcal{O}(0.1)$ are obtained from~\cite{Cirelli:2013ufw}. The function $G_{\phi}(x)$ is given by~\cite{Ishizuka:1990}:
\begin{align}
G_{\phi}(x) &= 1 - \frac{5}{2}x^2 - \frac{35}{22}x^4 + \frac{5}{64} \left( 28x^3 + 5x^5 \right) \arctan\left(\frac{2}{x}\right) \nonumber \\
&\quad + \frac{5}{64} \frac{\sqrt{2}x^6}{\sqrt{2+x^2}} \arctan\left(\frac{2\sqrt{2(2+x^2)}}{x^2}\right).
\end{align}
The additional factor $\xi(T,m_\phi)$ accounting for the finite scalar mass effect is defined as~\cite{Krnjaic:2015mbs}:
\begin{align}
\xi(T,m_\phi) &\equiv \frac{\int_{m_\phi}^\infty dx\frac{x\sqrt{x^2-m_\phi^2}}{e^{x/T}-1}}{\int_0^\infty dx \frac{x^2}{e^{x/T}-1}}.
\end{align}

The total energy loss rate is thus given as 
\begin{align}
    \mathcal{P} = P_{\rm esc}Q_\phi V_{\rm SN}
\end{align}
where $Q_\phi$ is given above as the energy rate carried by the $\phi$ production, $V_{\rm SN}=4/3\pi R_{\rm SN}^3$ with $R_{\rm SN}\approx 10\,\rm km$ is the volume. $P_{\rm esc}$ represents the escape rate of $\phi$, which is given as~\cite{Krnjaic:2015mbs}:
\begin{align}
P_{\mathrm{esc}} &= \exp\left(-\frac{R_{\mathrm{SN}}}{\gamma c \tau_\phi}\right) \exp\left(-\frac{R_{\mathrm{SN}}}{\lambda_\phi}\right),
\end{align}
where $\gamma \tau_\phi$ is the average boosted lifetime of the scalar and $\lambda_\phi$ is its mean free path. The mean free path of the scalar is estimated using detailed balance~\cite{Turner:1988}: $\lambda_\phi \sim \rho_\phi/Q_\phi$, where $\rho_\phi$ is the scalar energy density estimated in the equilibrium limit (at the temperature of the SN). Requiring $\mathcal{P}\lesssim10^{53}\,{\rm erg/s}$ imposes stringent limit on the mass and couplings of the scalar.

\subsubsection*{Meson Decays}

The light scalars in BSM can also contribute to meson decays, including $B\to KH_5^0$ and $K\to \pi H_5^0$ combining with either $H_5^0\to\ell\ell$ or $H_5^0\to\nu\nu$. The decay width for $B\to KH_5^0$ and $K\to \pi H_5^0$ have been listed in~\autoref{sec:H5Production}. There have been several experiments searching for BSM signals through these channels. The LHCb collaboration provides the 95\% CL upper limits on ${\rm Br}(B\to K \phi)\times {\rm Br}(\phi\to \mu\mu)$ as functions of the scalar mass and its lifetime~\cite{LHCb:2015nkv,LHCb:2016awg}. The NA62 experiment at CERN has analyzed the rare decay $K^+\to\pi^+\nu\bar{\nu}$ and established model-independent upper limits on the branching ratio ${\rm Br}(K^+\to\pi^+\phi)$ for various mass $m_\phi$. The results are provided for stable or invisibly decaying particles, as well as the visible decays into SM particles with limits depending on the mass $m_\phi$ and lifetime $\tau_\phi$ of the scalar~\cite{NA62:2021zjw}. MicroBooNE experiment also provides a model-independent upper bounds on ${\rm Br}(K^+\to\pi^+\phi)\times{\rm Br}(\phi\to e^+e^-)$ at 95\% CL with the dependence on the mass and lifetime of the scalar~\cite{MicroBooNE:2021sov}. The E787/E949 experiment provides the 90\% CL upper limits on the branching ratio ${\rm Br}(K^+\to \pi^+\phi)$ where $\phi$ should escape the detector or decays invisibly as function of the mass and the lifetime of $\phi$~\cite{BNL-E949:2009dza}.

\subsubsection*{LEP}

The LEP experiments have also searched for light scalar signals through $e^+e^-\to Z^*\phi$ channel with about $3\times10^6$ hadronic Z decays being detected. The scalar $\phi$ can be either stable/long-lived or prompt decayed. The L3, ALEPH and OPAL collaborations have provided the upper bound on the production cross section or the coupling strength with the $Z$ boson at 95\% CL~\cite{L3:1996ome,ALEPH:1993sjl,OPAL:2007qwz}. All these constraints can be converted accordingly for $H_5^0$ in GM model.

\subsection{Forward Detectors}

\begin{figure}[!tbp]
    \centering
    \includegraphics[width=\textwidth]{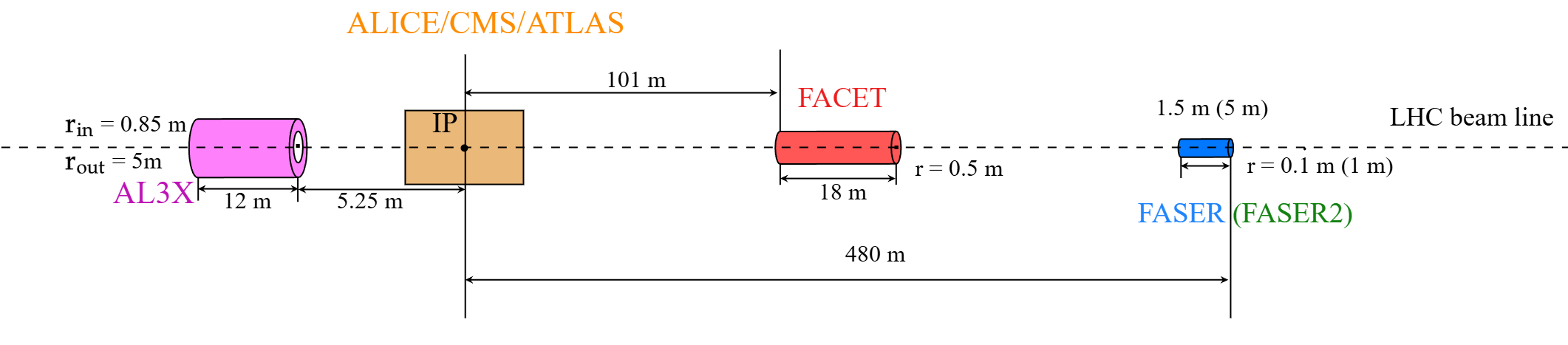}
    \caption{The position of forward detectors(AL3X, FACET, FASER/FASER2) relative to ALICE/CMS/ATLAS interaction point(IP). }	\label{fig:Forward_detectors}
\end{figure}

AL3X~\cite{Gligorov:2018vkc}, FACET~\cite{Cerci:2021nlb}, FASER/FASER2~\cite{FASER:2018eoc} are forward detectors located in the forward region within a range of several meters to several hundred meters from the interaction point(IP) at LHC. The position relative to the IP is shown in~\autoref{fig:Forward_detectors}. Specifically,
\begin{itemize}
    \item AL3X~\cite{Gligorov:2018vkc} is a cylindrical detector located near ALICE at a distance of 5.25 m from the IP, with an inner radius of 0.85 m, an outer radius of 5 m, and a length of 12 m. In the era of high-luminosity, the integrated luminosity of AL3X can reach 250 $\rm{fb}^{-1}$.
    \item FACET~\cite{Cerci:2021nlb} is a cylindrical detector located at a distance of 119 m from the IP near CMS, with a radius of 0.5 m and a length of 18 m. At HL-LHC, the integrated luminosity corresponding to FACET is 3000 $\rm{fb}^{-1}$.
    \item FASER/FASER2~\cite{FASER:2018eoc} is a cylindrical detector located near ATLAS at a distance of 480 m from the IP, with a radius of 0.1/1 m and a length of 1.5/5 m and luminosity of $150/3000\,\rm fb^{-1}$.
\end{itemize}

Based on the position and size of the various forward detectors (AL3X, FACET, FASER/FASER2), the probability of light CP-even fiveplet $H_5^0$ decaying in each detector, in the lab frame, is given as
\begin{align}
\mathcal{P}=\exp\left[-\frac{\ell_{1}}{\gamma\beta\tau}\right]-\exp\left[-\frac{\ell_{2}}{\gamma\beta\tau}\right]
\label{equ:decay_proability}
\end{align}
where $\beta$ is the velocity of a particle, $\gamma=(1-\beta^2)^{-1/2} $ is the Lorentz factor, $\tau$ is the proper lifetime of $H_5^{0}$, $\ell_1$ and $\ell_2$ are the distances between the boundaries of the forward detector and the IP along the direction of the momentum. Then the signal event that can be observed within the far detectors for given model parameters can be obtained as
\begin{align}
    S = \mathcal{L}\times\sum_i \sigma_i\times\mathcal{P}(p_i, \theta_i),
\end{align}
where we have summed over all bins as in~\autoref{fig:gm_scalar_rate}, $\sigma_i$ is the light scalar production cross section in each bin. The velocity and Lorentz boost factor $\beta,\gamma$ can be obtained from $p_i$ and the scalar mass, while $\theta_i$ and the configuration of the detectors determine $\ell_1$ and $\ell_2$ along the flying direction. Such searches can be safely treated background free. Hence, the exclusion region at 95\% CL is given by $S>3$.

\subsection{Results for the light CP-even fiveplet $H_5^0$}

\begin{figure}[!tbp]
	\centering{\includegraphics[width=0.49\textwidth]{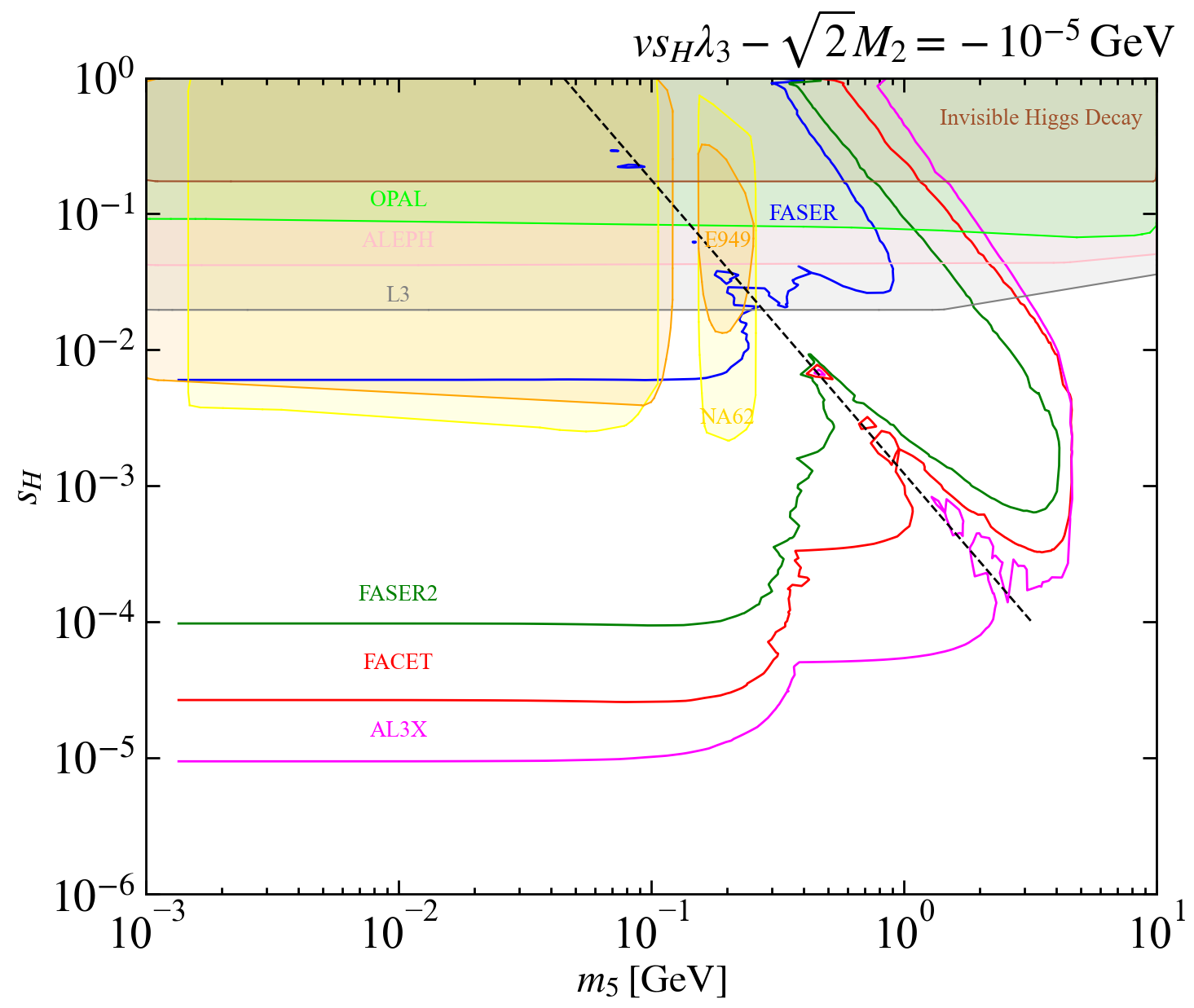}}\put(-185,30){$\rm{\MakeUppercase{\romannumeral 1}}$}
    \centering{\includegraphics[width=0.49\textwidth]{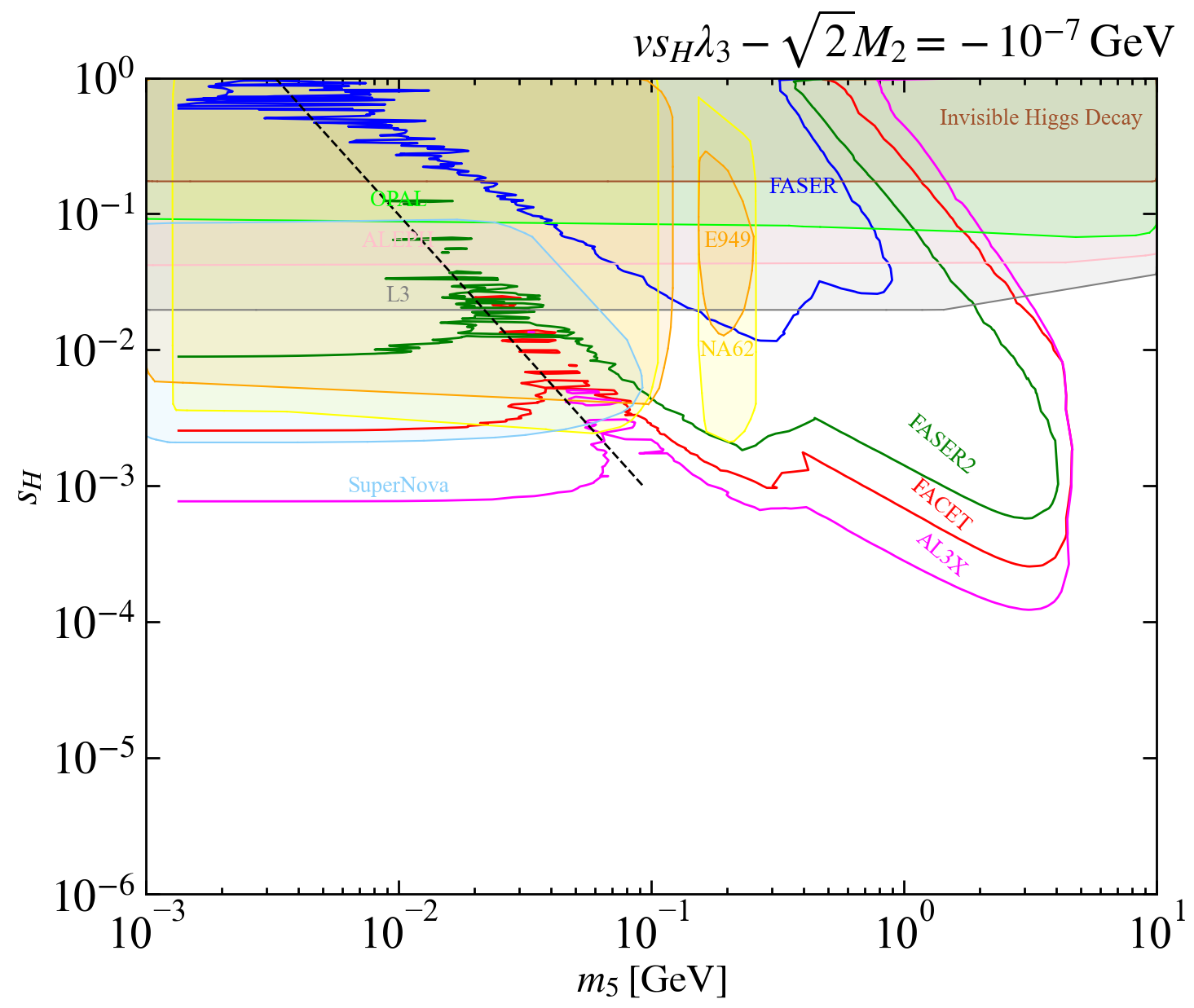}}\put(-185,30){$\rm{\MakeUppercase{\romannumeral 2}}$}\\
	\centering{\includegraphics[width=0.49\textwidth]{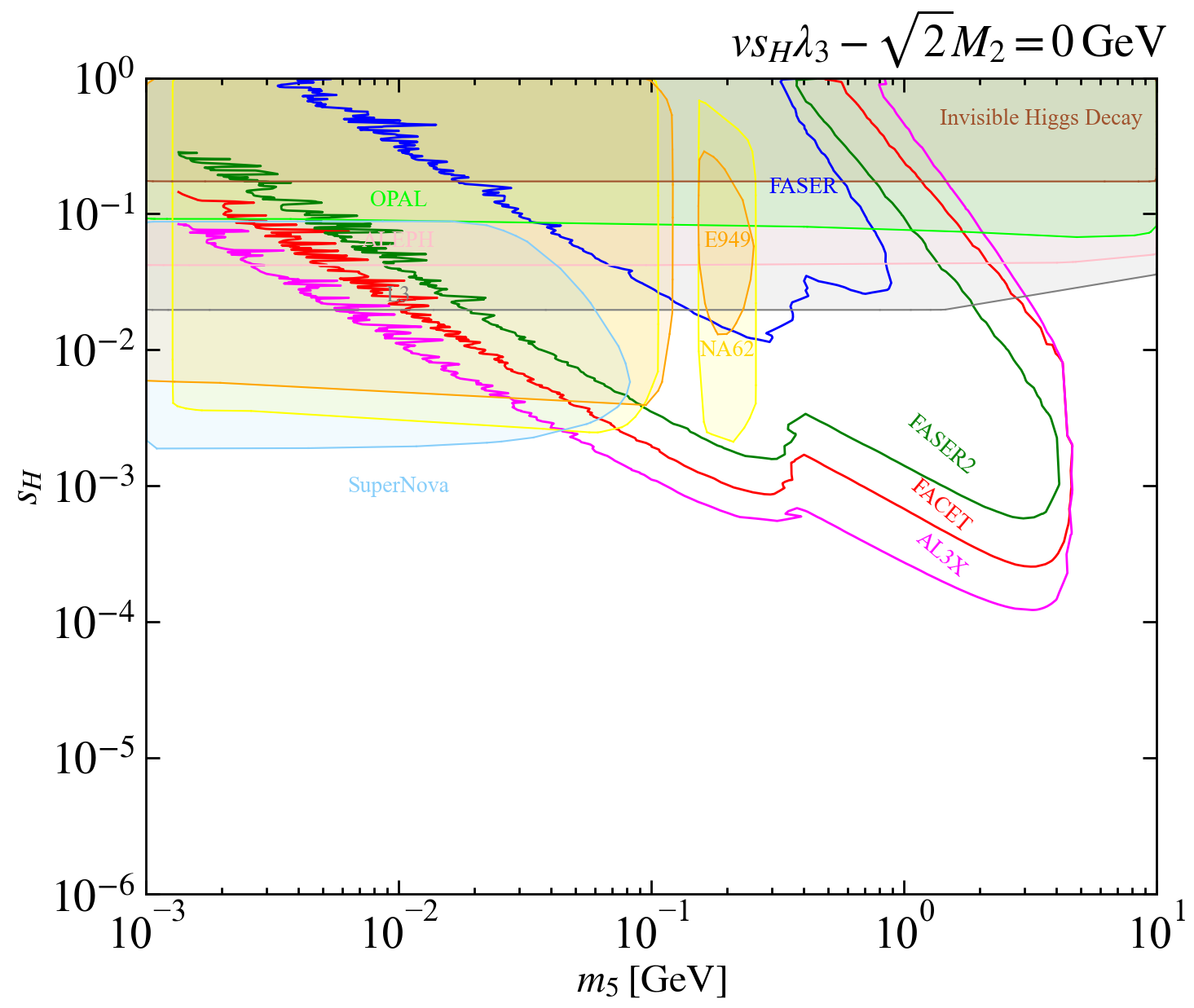}}\put(-185,30){$\rm{\MakeUppercase{\romannumeral 3}}$}
    \centering{\includegraphics[width=0.49\textwidth]{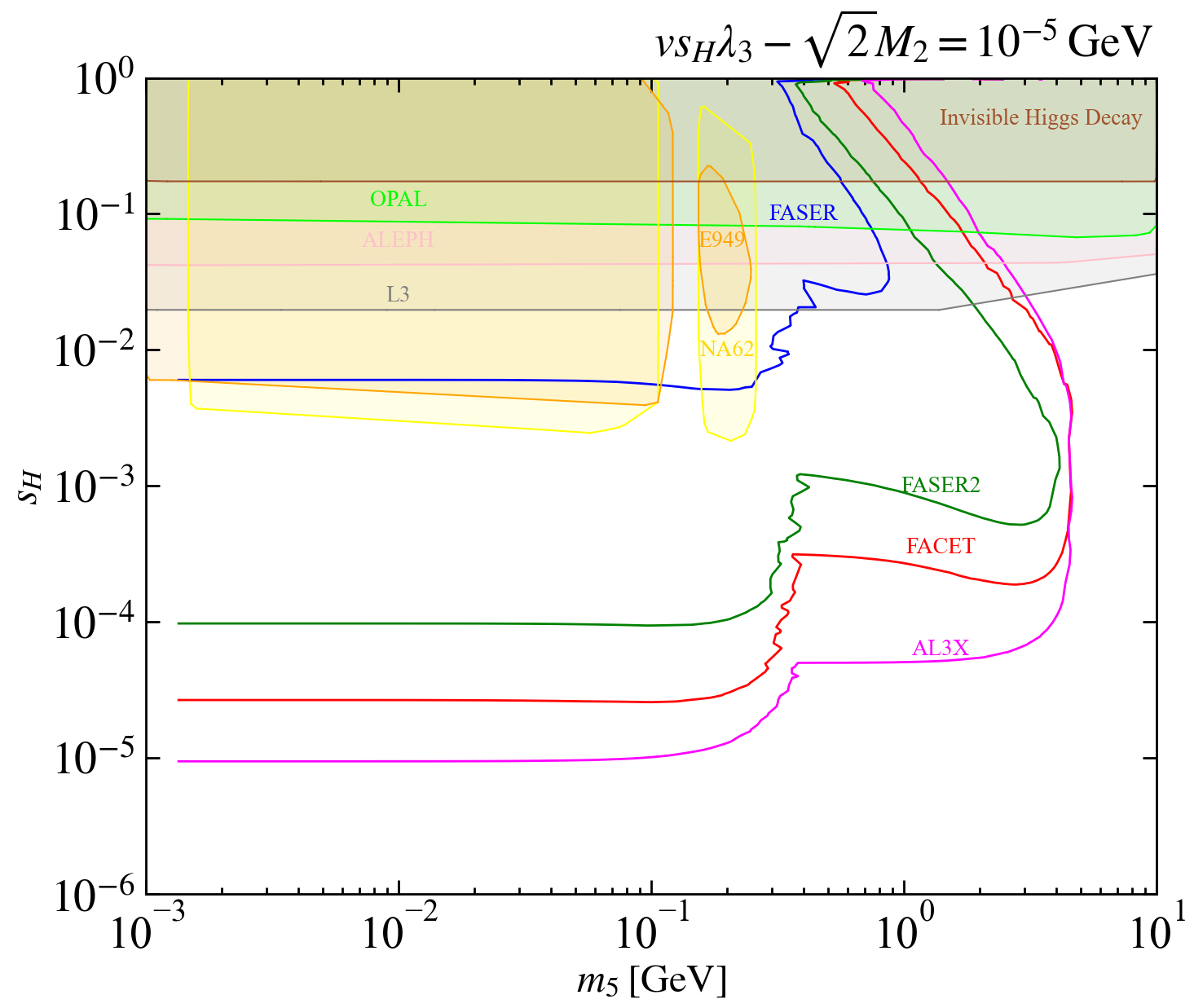}}\put(-185,30){$\rm{\MakeUppercase{\romannumeral 4}}$}
	\caption{Experimental exclusion limits at the 95\% confidence level for the light CP-even fiveplet $H_5^0$, including constraints from four forward detectors (AL3X, FACET, FASER, FASER2) and current experimental bounds (supernova, B meson decay, Kaon meson decay, and $e^+e^-\rightarrow Z^* \phi$ at LEP.}
	\label{fig:DDC_result}
\end{figure}

In~\autoref{fig:DDC_result}, we present the exclusion region at 95\% CL for light CP-even scalar $H_5^0$ from forward detectors on $m_5$-$s_H$ plane as indicated by the solid lines for four different benchmarks of $vs_H\lambda_3-\sqrt{2}M_2$ for AL3X, FACET and FASER/FASER2. The difference for different detectors mainly comes from the solid angle coverage. The exclusion regions from other experiments are also shown as shaded regions. The constraints from terrestrial experiments including Higgs invisible decay, LEP experiments, E949 and NA62 experiments are almost independent of $vs_H\lambda_3-\sqrt{2}M_2$. However, the supernova constraint varies according to $vs_H\lambda_3-\sqrt{2}M_2$ and appears only when $|vs_H\lambda_3-\sqrt{2}M_2|$ is sufficiently small. All these constraints can cover $s_H\gtrsim10^{-3}/10^{-2}$ for MeV/GeV scale scalar. On the other hand, the forward detectors can be complementary to current searches around sub-GeV region with small $|vs_H\lambda_3-\sqrt{2}M_2|$.

For forward detectors, the results are affected mainly by two factors: the production from meson decay and the total decay width of $H_5^0$. From the above discussion about the $H_5^0$ production, it is clear that for $m_5$ below several GeV, the cross section has minor dependence on $m_5$ but proportional to $s_H^2$. Hence the flux of $H_5^0$ at forward region is heavily suppressed by $s_H^2$. When the mass of the scalar is above several GeV, the production cross section is heavily suppressed. Consequently the exclusion regions are cut off around 4 GeV. The decay of $H_5^0$ is dominated by $H_5^0\to\gamma\gamma$ which, as we have discussed above, has two main contributions ($W$ contribution and scalar contribution). At lower left region in the $m_5$-$s_H$ plane, the scalar contribution is dominant, while the upper right region is dominated by $W$ contribution. The boundary in $m_5$-$s_H$ plane is determined by $vs_H\lambda_3-\sqrt{2}M_2$. When $vs_H\lambda_3-\sqrt{2}M_2<0$ these two contributions have strong cancellation around the boundary. With $vs_H\lambda_3-\sqrt{2}M_2=0$, only $W$ contribution remains which is the case of benchmark-III.

In the upper right region of $m_5$-$s_H$ plane with $m_5$ less than several GeV, the production cross section is sufficient. Hence, in this region, the exclusion lines are aligned with the decay width contours and is dominated by the $W$ contribution. At leading order, $\Gamma\sim s_H^2\frac{m_5^3}{v^2}$ for $W$ contribution. Hence the slope for the exclusion lines in this region is about $-3/2$. On the other hand, in the lower left region, the scalar contribution dominates. With $vs_H\lambda_3-\sqrt{2}M_2$ fixed for each panel, the exclusion lines are mainly determined according to the cross section. The only exception is benchmark-III where only $W$ contribution remains, the exclusion lines still have strong dependence on $s_H$. Further, in the upper panels, where $vs_H\lambda_3-\sqrt{2}M_2<0$, we have the cancellations between $W$ and scalar contributions to the scalar decay. The region where the cancellation happens is indicated by the dashed lines where we lost the sensitivity due to the extremely large decay length within this region.

\section{Summary}
\label{Sec:Conclusion}

The current searches for the new particles at the LHC have put stringent constraints on the parameter space of BSM models imposing strong upper bounds on the couplings of these possible new states with the SM particles. In such case, these particles might become long-lived. The searches for long-lived particles have been very active recently. In this work, we consider the searches for the CP-even light fermiophobic scalar $H_5^0$, which becomes long-lived at low-mass region, at the foward physics facilities including AL3X, FACET and FASER/FASER2.

Focusing on the fermiophobic scalar $H_5^0$ in the GM model, we obtained the loop-induced couplings between the scalar and fermions in the SM. However, the calculation is performed as general as possible. The results can be adapted to any fermiophobic scalar other than the fiveplet in GM model. 
With these loop-induced couplings between $H_5^0$ and the SM fermions, $H_5^0$ can be produced from the meson decays at low mass region.
The production of $H_5^0$ at sub-GeV region is almost independent of its mass. At higher mass, as most meson decay channels are closed due to the phase space, the production cross section drops dramatically when the mass is above several GeV. The production rate also proportional to $s_H^2$ which indicates the fraction of the $SU(2)_L$ triplet vev as it controls all the couplings of the $SU(2)_L$ triplet scalars with the SM particles. Although, the production will be heavily suppressed at low $s_H$ region which is however compensated by the large rate of the light meson productions in the forward region. Hence, in this region, sufficient flux of $H_5^0$ can be produced.

On the other hand, the same loop induced couplings also dominate the decays of $H_5^0$ in low mass region. The leading channel is the diphoton decay $H_5^0\to\gamma\gamma$ which is several magnitude larger than other ferminoic channels. The possible cancellation between $W$ and scalar contributions is identified. 
Other than the dependence on $m_5$, the $W$ contribution is proportional to $s_H$, while the scalar contributions are proportional to $vs_H\lambda_3-\sqrt{2}M_2$. Hence the cancellation and the boundary between regions with different contribution dominant can be determined by $s_H$ and $m_5$ with fixed $vs_H\lambda_3-\sqrt{2}M_2$.
The decay width behaves differently in different region where either $W$ or scalar contributions dominate. 

There are also many other searches that are sensitive to such light scalar. We considered the constraints from Higgs invisible decay, beam dump experiments, the searches at the LEP as well as from supernova. The terrestrial experiments are robust. They mainly exclude the region with large $s_H$ which determines the couplings between the fiveplet with the SM particles. The constraint from supernova can be strong, but only within a limited parameter space.

The searches from forward detectors extend the sensitivity to even smaller $s_H$. However, the exclusion region will change according to the choice of the parameter space and is mainly determined by the dependence of production cross section and the decay width on the model parameters. At large $s_H$ region, the production cross section is sufficient, the exclusion lines are mainly determined by the decay. In this region, $W$ contribution to the scalar decay dominates in general. At leading order we have $\Gamma\sim s_H^2\frac{m_5^3}{v^2}$ from $W$ contribution. Consequently, the slope for the contours in this region is about $-3/2$. When $s_H$ is small, for a fixed but non-zero value of $vs_H\lambda_3-\sqrt{2}M_2$, the scalar contribution dominates for the decay width. Further, the production rate is also suppressed by $s_H^2$. In this case, the decay width has minor dependence on the scalar mass, the exclusion lines are mainly determined by the production rate. For $vs_H\lambda_3-\sqrt{2}M_2=0$, only $W$ contribution for the decay exists, the exclusion line is determined by the combined effects of production and decay. In the parameter regions we considered, we find that the searches at forward detectors can be more sensitive than the current constraints, especially for mass around GeV scale. 
Hence, the searches from forward detectors are complementary to the current searches, and will further explore the situation where the particle is long-lived and provide insight into the BSM models.

\acknowledgments
YF, XW, YW and YZ are supported by the National Natural Science Foundation of China (NNSFC) under grant No.12305112. WS is supported by NNSFC under grant No.12305115. The authors gratefully acknowledge the valuable discussions and insights provided by the members of the China Collaboration of Precision Testing and New Physics (CPTNP).

\appendix
\section{The couplings relevant for the loop induced FCNC couplings of $H_5^0$}
\label{app:couplings}
\subsubsection*{\texorpdfstring{$S_1S_2S_3$}{S1S2S3}}
The couplings involving three scalars are listed below:
\begin{subequations}
\begin{align}
    g_{G^0G^0H_5^0} &= -2g_{G^+G^-H_5^0} = \frac{2\sqrt{3}}{3}s_H^3\lambda_3v + \sqrt{3}c_H^2s_H\lambda_5v + \sqrt{\frac{2}{3}}c_H^2 M_1 + 2\sqrt{6}s_H^2M_2 \\
    g_{G^0H_3^0H_5^0} &= -2g_{G^+H_3^-H_5^0} = -2g_{H_3^+G^-H_5^0} \nonumber \\
    &= \frac{2\sqrt{3}}{3}c_Hs_H^2\lambda_3v - \frac{\sqrt{3}}{3}c_H(1-3c_H^2)\lambda_5v - \sqrt{\frac{2}{3}}c_Hs_HM_1 + 2\sqrt{6}c_Hs_HM_2\\
    g_{H_3^0H_3^0H_5^0} &= -2g_{H_3^+H_3^-H_5^0} \nonumber \\
    & = \frac{2\sqrt{3}}{3}c_H^2s_H\lambda_3v - \frac{\sqrt{3}}{3}s_H(1+3c_H^2)\lambda_5v + \sqrt{\frac{2}{3}}s_H^2M_1 + 2\sqrt{6}c_H^2M_2
\end{align}
\end{subequations}

\subsubsection*{\texorpdfstring{$\bar{F}_1F_2S$}{F1F2S}}
The Yukawa couplings involved in the calculation are listed below, where for quarks the color indices are omitted.
\begin{subequations}
\begin{align}
(g_{\bar{d}_id_jG^0}^L, g_{\bar{d}_id_jG^0}^R) &= i\delta_{ij}c_H \frac{y_{d_i}}{\sqrt{2}}(1, -1)\\
(g_{\bar{u}_iu_jG^0}^L, g_{\bar{u}_iu_jG^0}^R) &= -i\delta_{ij}c_H \frac{y_{u_i}}{\sqrt{2}}(1, -1)\\
(g_{\ell^+\ell^-G^0}^L, g_{\ell^+\ell^-G^0}^R) &= i\delta_{ij}c_H \frac{y_{\ell_i}}{\sqrt{2}}(1, -1)\\
(g_{\bar{d}_id_jH_3^0}^L, g_{\bar{d}_id_jH_3^0}^R) &= -i\delta_{ij}s_H \frac{y_{d_i}}{\sqrt{2}}(1, -1)\\
(g_{\bar{u}_iu_jH_3^0}^L, g_{\bar{u}_iu_jH_3^0}^R) &= i\delta_{ij}s_H \frac{y_{u_i}}{\sqrt{2}}(1, -1)\\
(g_{\ell^+\ell^-H_3^0}^L, g_{\ell^+\ell^-H_3^0}^R) &= -i\delta_{ij}s_H \frac{y_{\ell_i}}{\sqrt{2}}(1, -1)\\
(g_{\bar{u}_id_jG^+}^L, g_{\bar{u}_id_jG^+}^R) &= c_H V_{ij}(y_{u_i}, -y_{d_j})\\
(g_{\bar{d}_iu_jG^-}^L, g_{\bar{d}_iu_jG^-}^R) &= -c_H (V_{ji})^* (y_{d_i}, -y_{u_j})\\
(g_{\bar{\nu}_i\ell_j G^+}^L, g_{\bar{\nu}_i\ell_j G^+}^R) &= -\delta_{ij}c_H y_{\ell_j} (0, 1)\\
(g_{\bar{\ell}_i\nu_j G^-}^L, g_{\bar{\ell}_i\nu_j G^-}^R) &= -\delta_{ij}c_H y_{\ell_i} (1, 0)\\
(g_{\bar{u}_id_jH_3^+}^L, g_{\bar{u}_id_jH_3^+}^R) &= -s_H V_ {ij} (y_{u_i}, -y_{d_j})\\
(g_{\bar{d}_iu_jH_3^-}^L, g_{\bar{d}_iu_jH_3^-}^R) &= s_H (V_{ji})^* (y_{d_i}, -y_{u_j})\\
(g_{\bar{\nu}_i\ell_j H_3^+}^L, g_{\bar{\nu}_i\ell_j H_3^+}^R) &= \delta_{ij}s_H y_{\ell_j} (0,1)\\
(g_{\bar{\ell}_i\nu_j H_3^-}^L, g_{\bar{\ell}_i\nu_j H_3^-}^R) &= \delta_{ij}s_H y_{\ell_i} (1,0)
\end{align}
\end{subequations}

\subsubsection*{\texorpdfstring{$S_1S_2V$}{S1S2V}}

\begin{subequations}
\begin{align}
    (g_{G^0H_5^0Z}^{G^0}, g_{G^0H_5^0Z}^{H_5^0}) &= i\frac{gs_H}{\sqrt{3}c_W} (1,-1)\\
    (g_{H_3^0H_5^0Z}^{H_3^0}, g_{H_3^0H_5^0Z}^{H_5^0}) &= i\frac{gc_H}{\sqrt{3}c_W}(1,-1)\\
    (g_{G^\pm H_5^0W^\mp}^{G^\pm}, g_{G^\pm H_5^0W^\mp}^{H_5^0}) &= \pm\frac{gs_H}{2\sqrt{3}}(1, -1)\\
    (g_{H_3^\pm H_5^0W^\mp}^{H_3^\pm}, g_{H_3^\pm H_5^0W^\mp}^{H_5^0}) &= \pm\frac{gc_H}{2\sqrt{3}}(1, -1)
\end{align}
\end{subequations}

\subsubsection*{\texorpdfstring{$\bar{F}_1F_2V$}{F1F2V}}

\begin{subequations}
\begin{align}
(g_{\bar{d}_id_jZ}^L, g_{\bar{d}_id_jZ}^R) &= \delta_{ij} \frac{g}{c_W}\left( - \frac{1}{2}+\frac{1}{3}s_W^2, \frac{1}{3}s_W^2\right) \\
(g_{\bar{u}_iu_jZ}^L, g_{\bar{u}_iu_jZ}^R) &= \delta_{ij} \frac{g}{c_W}\left(\frac{1}{2}-\frac{2}{3}s_W^2, - \frac{2}{3}s_W^2 \right) \\
(g_{\bar{\ell}_i\ell_jZ}^L,g_{\bar{\ell}_i\ell_jZ}^R) &= \delta_{ij} \frac{g}{c_W}\left(-\frac{1}{2}+s_W^2, s_W^2\right)\\
(g_{\bar{\nu}_i\nu_jZ}^L, g_{\bar{\nu}_i\nu_jZ}^R) &= \delta_{ij} \frac{g}{c_W} \frac{1}{2}(1,0)\\
(g_{\bar{d}_iu_jW^-}^L, g_{\bar{d}_iu_jW^-}^R) &= \frac{g}{\sqrt{2}}(V_{ji})^*(1,0)\\
(g_{\bar{u}_id_jW^+}^L, g_{\bar{u}_id_jW^+}^R) &= \frac{g}{\sqrt{2}}V_{ij}(1,0)\\
(g_{\bar{\ell}_i\nu_jW^-}^L, g_{\bar{\ell}_i\nu_jW^-}^R) &= \delta_{ij} \frac{g}{\sqrt{2}}(1,0)\\
(g_{\bar{\nu}_i\ell_jW^+}^L, g_{\bar{\nu}_i\ell_jW^+}^R) &= \delta_{ij} \frac{g}{\sqrt{2}} (1,0)
\end{align}
\end{subequations}

\subsubsection*{\texorpdfstring{$V_1V_2S$}{V1V2S}}

\begin{subequations}
\begin{align}
g_{ZZH_5^0} &= -\frac{g^2 s_H v}{\sqrt{3}c_W^2} \\
g_{W^+W^-H_5^0} &=  \frac{g^2 s_H v}{2\sqrt{3}}
\end{align}
\end{subequations}

\bibliographystyle{bibsty}
\bibliography{references}

\end{document}